\documentstyle[preprint,eqsecnum,amssymb,aps]{revtex}

\begin{document}

\preprint{quant-ph/9504008}
\title{Photon states associated with Holstein-Primakoff \\
realization of SU(1,1) Lie algebra}
\author{C. Brif \footnote{e-mail: phr65bc@phys1.technion.ac.il}}
\address{Department of Physics, Technion--Israel Institute of
 Technology, Haifa 32000, Israel}
\maketitle

\begin{abstract}

Statistical and phase properties and number-phase uncertainty
relations are systematically investigated for photon states
associated with the Holstein-Primakoff realization of the SU(1,1)
Lie algebra. Perelomov's SU(1,1) coherent states and the eigenstates
of the SU(1,1) lowering generator (the Barut-Girardello states) are
discussed. A recently developed formalism, based on the antinormal
ordering of exponential phase operators, is used for studying phase
properties and number-phase uncertainty relations. This study shows
essential differences between properties of the Barut-Girardello
states and the SU(1,1) coherent states.
The philophase states, defined as states with simple phase-state
representations, relate the quantum description of the optical phase
to the properties of the SU(1,1) Lie group.
A modified Holstein-Primakoff realization is derived, and eigenstates
of the corresponding lowering generator are discussed. These states
are shown to contract, in a proper limit, to the familiar Glauber
coherent states.

\end{abstract}

\pagebreak

\section{Introduction}
\label{sec:intro}

The quantum mechanical calculations are based on appropriate sets of
states in Hilbert space of a system. The basic system of quantum
electrodynamics is the quantized single-mode electromagnetic field
which is modeled by the quantum harmonic oscillator.
The corresponding basic
set of states is the complete orthonormal set of the number states
$| n \rangle$ $(n=0,1,\ldots ,\infty)$, that can be used for
expanding all the field states. From the other hand, the overcomplete
set of the Glauber coherent states (CS) $| \alpha \rangle$
\cite{Gla,Sud} has a number of remarkable
properties and has been proved to be extremely useful in quantum
optics \cite{Gla,KlSud}. The Glauber CS are closely associated with
the boson creation and annihilation operators $\hat{a}^{\dagger}$
and $\hat{a}$, which, together
with the identity operator $\hat{1}$, form a realization of the
Heisenberg-Weyl nilpotent Lie algebra \cite{Weyl}. The corresponding
Lie group is the dynamical symmetry group of Hamiltonians for a
number of important quantum mechanical problems \cite{Per77,Per86}.
The Glauber CS $| \alpha \rangle$
can be defined in three ways \cite{Gla,Gil_Rev}:
(a) the eigenstates of the lowering
operator $\hat{a}$; (b) minimum-uncertainty states or, more
generally, intelligent states for position and momentum;
(c) states constructed by action of displacement operators, which
represent group elements, on the vacuum state. For the
Heisenberg-Weyl group all these definitions are equivalent, but for
other Lie groups they lead to distinct states.

In the present work we concentrate on the SU(1,1) Lie group whose
algebra has a number of realizations related to the quantized light
field. The most known of them are the single-mode realization in
terms of the amplitude-squared boson operators and the two-mode
realization in terms of the boson creation and annihilation operators
of the modes \cite{Per77,Per86}. These realizations and states
associated with them have been studied in connection with the field
quadrature squeezing \cite{WodEb,Ger_PR,Schum}.
We consider here the Holstein-Primakoff (HP) single-mode realization
of the SU(1,1) Lie algebra \cite{HP,Ger_JP}. The analogous
realization had been formerly introduced by Holstein and Primakoff
\cite{HP} for SU(2), and in the SU(1,1) case it is given by
\cite{Ger_JP}
	\begin{equation}
\hat{K}_{+}(k) = \sqrt{\hat{a}^{\dagger}\hat{a} + 2k - 1}\,
\hat{a}^{\dagger} , \;\;\;\;
\hat{K}_{-}(k) = \hat{a} \sqrt{\hat{a}^{\dagger}\hat{a} + 2k - 1} ,
\;\;\;\;
\hat{K}_{3}(k) = \hat{a}^{\dagger}\hat{a} + k .
\label{1.1}
	\end{equation}
Here $k$ is the Bargmann index labeling unitary irreducible
representations of the SU(1,1) Lie group [see text after Eq.\
(\ref{2.5})]. Aharonov {\em et al.} \cite{Ahar} have written the
HP SU(1,1) realization in another form, by
using the number operator $\hat{n} = \hat{a}^{\dagger}\hat{a}$ and
the Susskind-Glogower exponential phase operators
$\widehat{e^{i\phi}}$ and $\widehat{e^{-i\phi}}$ \cite{SG,CN}.
It was shown \cite{Ahar,Lern,LL,Vou90,Vou92,Vou93} that the HP
realization is related to the problem of the quantum description of
optical phase. The generators (\ref{1.1})
of the HP realization with $k=\frac{1}{2}$ have been used in the
Jaynes-Cummings model Hamiltonians with intensity-dependent
coupling \cite{JCM}. The SU(1,1) Lie group is the dynamical symmetry
group of these Hamiltonians. A multiboson version of the HP
realization (\ref{1.1}) have been constructed by using generalized
boson operators \cite{Katr}.

Various states associated with the HP SU(1,1) realization exist in
the harmonic oscillator Hilbert space. These states can be
conveniently treated by using general group-theoretical techniques.
One can consider the generalized CS obtained by action of SU(1,1)
group elements on the vacuum state
\cite{Per72,Per77,Per86,Vou92,Vou93}, and the eigenstates of the
lowering generator $\hat{K}_{-}(k)$ (the so-called Barut-Girardello
states \cite{BG,BBA_QO}).
The present paper is devoted to the systematic investigation of
properties of these states.
It is known that the SU(1,1) CS are closely related to phase states
\cite{Ahar,Lern,LL,Vou90,Vou92,Vou93}.
We proceed with a subsequent development of this relation by
constructing a class of states characterized by simple phase-state
representations \cite{BBA_PR2}. An especial attention is devoted to
the study of phase properties and uncertainty relations between the
number and phase observables.
We start in Sec.\ \ref{sec:cs} with the SU(1,1) CS. Phase properties
and number-phase uncertainty relations are examined using a recently
developed formalism, based on the antinormal ordering of the
Susskind-Glogower exponential phase operators \cite{LuPer,BBA_PR1}.
In Sec.\ \ref{sec:bg} we discuss in detail statistical
and phase properties of the Barut-Girardello states \cite{BG,BBA_QO}.
These states have sub-Poissonian photon statistics and form an
overcomplete basis in the harmonic oscillator Hilbert space. This
basis was used to construct a diagonal representation of the density
operator, which was shown to be well-behaved for nonclassical
photon states \cite{BBA_QO}. A class of generalized {\em philophase
states\/} (states with simple phase-state representations
\cite{BBA_PR2}) is considered in Sec.\ \ref{sec:pp}, and states
with sub-Poissonian statistical properties are found.
A modified HP SU(1,1) realization is obtained by using the
antinormal ordering of the exponential phase operators.
Eigenstates of the modified lowering generator can be described
as philophase states in one limit and contract to the Glauber CS
in another.
In Sec.\ \ref{sec:conc} we summarize our conclusions.

\section{The SU(1,1) coherent states}
\label{sec:cs}

The concept of CS has been generalized by Perelomov
\cite{Per72,Per77,Per86} and Gilmore \cite{Gil,Gil_Rev}
for arbitrary Lie group. Generalized CS are obtained by action
of group elements on an extreme state of the group Hilbert space and
thus can be created by using Hamiltonians for which given Lie group
is the dynamical symmetry group. We start our discussion by a brief
review of basic properties of SU(1,1) Lie group.
For a given value of $k$, the operators (\ref{1.1}) form a
realization of the SU(1,1) Lie algebra \cite{SU11},
	\begin{equation}
[\hat{K}_{3}(k) , \hat{K}_{\pm}(k)] = \pm \hat{K}_{\pm}(k) ,
\;\;\;\;\; [\hat{K}_{-}(k) , \hat{K}_{+}(k)] = 2\hat{K}_{3}(k) .
\label{2.2}
	\end{equation}
The group SU(1,1), whose generators are  $\hat{K}_{\pm}(k)$ and
$\hat{K}_{3}(k)$, is the
most elementary noncompact non-Abelian simple Lie group. It has
several series of unitary irreducible representations: discrete,
continuous and supplementary \cite{SU11}. In the present work we
discuss only the case of the discrete series. The Casimir operator
$\hat{Q}(k)$ for any irreducible representation is the
identity times a number,
	\begin{equation}
\hat{Q}(k)= [\hat{K}_{3}(k)]^{2} -\frac{1}{2}
[\hat{K}_{+}(k)\hat{K}_{-}(k) +\hat{K}_{-}(k)\hat{K}_{+}(k)]
= k(k-1)\hat{1} .
\label{2.5}
	\end{equation}
Thus a representation of SU(1,1) is determined by a single number
$k$ (the Bargmann index); for the discrete series this number
acquires discrete values $k=\frac{1}{2},1,\frac{3}{2},2,\ldots$.
The corresponding state space is spanned by the complete orthonormal
basis $| k,n \rangle$ $(n=0,1,\ldots ,\infty)$,
	\begin{equation}
 \langle k,n|k,n' \rangle = \delta_{nn'} , \;\;\;\;\;
 \sum_{n = 0}^{\infty} | k,n \rangle \langle k,n | = \hat{1} .
 \label{2.6}
	\end{equation}
These states may be defined by the following relations \cite{SU11}
	\begin{equation}
 \begin{array}{rcl}
 \hat{K}_{3}(k) | k,n \rangle & = & (n+k) | k,n \rangle ,  \\
 \hat{K}_{+}(k) | k,n \rangle & = & \sqrt{(n+1)(n+2k)}\,
|k,n+1\rangle ,  \\
 \hat{K}_{-}(k) | k,n \rangle & = & \sqrt{n(n+2k-1)}\,
|k,n-1\rangle .
\end{array}   \label{2.7}
	\end{equation}
By considering the action of the operators $\hat{K}_{\pm}(k)$ and
$\hat{K}_{3}(k)$ of the form (\ref{1.1}) on the number-state basis
$| n \rangle$ of the harmonic
oscillator Hilbert space, we see that this number-state basis
coincide with the discrete series SU(1,1) orthonormal basis
$| k,n \rangle$ for any
allowed value of $k$. It means that the discrete series SU(1,1) state
space is equivalent to the harmonic oscillator Hilbert space, when
using the HP realization. Speaking elsewhere in the
following about SU(1,1) we mean its HP realization defined by
(\ref{1.1}), with the orthonormal
basis being merely the number-state basis.

\subsection{Coherent-state basis and the analytic representation
in the unit disk}
\label{cs_B}

As an example of the use of the group-theoretical methods, we
consider the general results of Perelomov \cite{Per72,Per77,Per86}
for SU(1,1) discrete series CS.
A coherent state is specified by pseudo-Euclidian unit vector
$\bbox{\tau}$ of the form
	\begin{equation}
\bbox{\tau} = (\cosh \tau , \sinh \tau \cos \varphi ,
\sinh \tau \sin \varphi ) .
\label{2.8}
	\end{equation}
The CS $| k,\zeta \rangle$ are obtained by applying unitary operators
$\hat{D}(\xi)$ to the vacuum state,
	\begin{equation}
 | k,\zeta \rangle = \exp \left[ \xi \hat{K}_{+}(k) -\, \xi^{\ast}
\hat{K}_{-}(k) \right] | 0 \rangle
 = (1-|\zeta|^{2})^{k} \exp \left[ \zeta \hat{K}_{+}(k) \right]
| 0 \rangle ,
\label{2.9}
	\end{equation}
where $\xi=-\frac{\tau}{2} e^{-i\varphi}$ and $\zeta=-\tanh
\frac{\tau}{2}\, e^{-i\varphi}$, so $|\zeta|<1$. Expanding the
exponential
and using (\ref{2.7}), one gets the decomposition of the CS over the
number-state basis,
	\begin{equation}
 | k,\zeta \rangle = (1-|\zeta|^{2})^{k} \sum_{n = 0}^{\infty}
\left[\frac{\Gamma(n+2k)
}{n!\Gamma(2k)}\right]^{1/2}\!\zeta^{n} | n \rangle .
 \label{2.10}
	\end{equation}
The condition $|\zeta|<1$ means that the `phase space' of the
SU(1,1) CS is the interior of the unit disk. The CS are normalized
but not orthogonal to each other:
	\begin{equation}
 \langle k,\zeta_{1} | k,\zeta_{2} \rangle =
\frac{ (1-|\zeta_{1}|^{2})^{k}
 (1-|\zeta_{2}|^{2})^{k} }{ (1-\zeta_{1}^{\ast} \zeta_{2})^{2k} } .
\label{2.11}
	\end{equation}
The identity resolution is an important property of the CS:
	\begin{equation}
\int d\mu (k,\zeta)  | k,\zeta \rangle \langle k,\zeta |
= \hat{1},
\label{2.12}
	\end{equation}
where
	\begin{equation}
 d\mu (k,\zeta) = \frac{2k-1}{\pi} \frac{d^{2}\!
\zeta}{(1-|\zeta|^{2})^{2}}  ,
\label{2.13}
	\end{equation}
and for $k=\frac{1}{2}$ the limit $k\rightarrow \frac{1}{2}$ must be
taken after the integration is carried out in the general form. Thus
the SU(1,1) CS form an overcomplete basis.

One can can represent the harmonic oscillator Hilbert space as the
Hilbert space of entire functions $f(k,\zeta)$, which are analytic
in the unit disk. For a normalized photon state
	\begin{equation}
 | f \rangle = \sum_{n = 0}^{\infty} C_{n}(f) | n \rangle ,
\label{2.15}
	\end{equation}
we get
	\begin{equation}
f(k,\zeta) = (1-|\zeta|^{2})^{-k} \langle k,\zeta^{\ast} | f \rangle
= \sum_{n = 0}^{\infty} C_{n}(f) \left[\frac{\Gamma(n+2k)
}{n!\Gamma(2k)}\right]^{1/2}\!\zeta^{n}  ,
\label{2.16}
	\end{equation}
and this state can be represented in the coherent-state basis:
	\begin{equation}
 | f \rangle = \int d\mu (k,\zeta)  (1-|\zeta|^{2})^{k}
f(k,\zeta^{\ast}) | k,\zeta \rangle .  \label{2.17}
	\end{equation}
We will refer to such representations as the representations in the
unit disk.
The generators $\hat{K}_{\pm}(k)$ and $\hat{K}_{3}(k)$ act on the
Hilbert space of entire
functions $f(k,\zeta)$ as first-order differential operators:
	\begin{equation}
\hat{K}_{+}(k) =  \zeta^{2} \frac{d}{d\zeta} + 2k\zeta ,  \;\;\;\;\;
 \hat{K}_{-}(k) = \frac{d}{d\zeta}  ,  \;\;\;\;\;
 \hat{K}_{3}(k)  =  \zeta \frac{d}{d\zeta} + k  .   \label{2.22}
	\end{equation}

\subsection{Phase states and phase-like states}
\label{cs_C}

We discuss now the CS $| k,\zeta \rangle$ with $k=\frac{1}{2}$.
This case is
interesting by two reasons. Firstly, phenomenological Jaynes-Cummings
model Hamiltonians with intensity-dependent coupling have been
constructed \cite{JCM}, for which the SU(1,1) group in the HP
$k=\frac{1}{2}$ representation is the dynamical symmetry group.
In principle, the SU(1,1) CS with $k=\frac{1}{2}$ can be created by
applying such Hamiltonians to the vacuum state. Secondly, the SU(1,1)
CS with $k=\frac{1}{2}$ are closely related to the phase
states $|\theta \rangle$, given by \cite{CN}
	\begin{equation}
|\theta \rangle = \frac{1}{\sqrt{2\pi}} \sum_{n = 0}^{\infty}
e^{in\theta} | n \rangle .
\label{2.23}
	\end{equation}
For $k=\frac{1}{2}$, we can write, by using the number operator
$\hat{n}$ and the exponential phase operators $\widehat{e^{i\phi}}$
and $\widehat{e^{-i\phi}}$,
	\begin{equation}
\hat{K}_{+}(\mbox{{\small $\frac{1}{2}$}})=\hat{n}
\widehat{e^{-i\phi}} , \mbox{\hspace{1cm}}
\hat{K}_{-}(\mbox{{\small $\frac{1}{2}$}})=
\widehat{e^{i\phi}} \hat{n}  , \mbox{\hspace{1cm}}
\hat{K}_{3}(\mbox{{\small $\frac{1}{2}$}})=\hat{n} +
\frac{1}{2} .
\label{2.24}
	\end{equation}
It was noted \cite{Lern,Ahar,LL} that the CS
	\begin{equation}
|\mbox{{\small $\frac{1}{2}$}} ,\zeta \rangle =
\sqrt{1-|\zeta|^2}\, \sum_{n = 0}^{\infty} \zeta^{n} |n\rangle
\label{2.26}
	\end{equation}
are the eigenstates of the lowering exponential phase operator
$\widehat{e^{i\phi}}$,
	\begin{equation}
\widehat{e^{i\phi}} |\mbox{{\small $\frac{1}{2}$}} ,\zeta \rangle =
\zeta |\mbox{{\small $\frac{1}{2}$}} ,\zeta \rangle ,
\label{2.28}
	\end{equation}
just like the phase states,
	\begin{equation}
\widehat{e^{i\phi}} |\theta \rangle = e^{i\theta} |\theta \rangle .
\label{2.29}
	\end{equation}
Therefore we will refer to the states $|\frac{1}{2} ,\zeta \rangle$
as the {\em phase-like
states\/}. These states depend on two real parameters, $|\zeta|$ and
${\rm arg}\, \zeta$, while for characterization of phase only one
real periodic parameter $\theta$ is needed.
When $|\zeta| \rightarrow 1$, one gets
	\begin{equation}
|\theta \rangle = \frac{1}{\sqrt{2\pi}} \lim_{|\zeta| \rightarrow 1}
(1-|\zeta|^2)^{-1/2} |\mbox{{\small $\frac{1}{2}$}} ,\zeta \rangle .
\label{2.30}
	\end{equation}

Phase states $|\theta \rangle$ of Eq.\ (\ref{2.23}) resolve the
identity,
	\begin{equation}
\int_{\theta_{0}}^{\theta_{0}+ 2\pi } d\theta \,
|\theta\rangle\langle\theta| = \hat{1} ,
\label{2.31}
	\end{equation}
where $\theta_{0}$ is a reference phase. Therefore, for arbitrary
normalized state $|f\rangle$ of the form (\ref{2.15}), the
phase-state representation is given by \cite{BBA_PR2}
	\begin{equation}
|f\rangle =  \frac{1}{\sqrt{2\pi}}
\int_{\theta_{0}}^{\theta_{0}+ 2\pi } d\theta \, \Theta(f;\theta)
|\theta\rangle ,
\label{2.32}
	\end{equation}
where
	\begin{equation}
\Theta(f;\theta) =  \sqrt{2\pi}\, \langle\theta|f\rangle =
\sum_{n = 0}^{\infty} C_{n}(f) e^{-in\theta} .
\label{2.33}
	\end{equation}
The SU(1,1) generators act on the $\Theta(f;\theta)$ as first-order
differential operators:
	\begin{equation}
\hat{K}_{+}(\mbox{{\small $\frac{1}{2}$}})=e^{-i\theta}
+ie^{-i\theta}\frac{d}{d\theta} , \;\;\;\;\;\;
\hat{K}_{-}(\mbox{{\small $\frac{1}{2}$}}) =
ie^{i\theta}\frac{d}{d\theta}, \;\;\;\;\;\;
\hat{K}_{3}(\mbox{{\small $\frac{1}{2}$}}) =
i\frac{d}{d\theta} +\frac{1}{2} .
\label{2.34}
	\end{equation}
Thus the phase-state representation is a limiting case of
the SU(1,1) coherent-state
representation when $k=\frac{1}{2}$ and $|\zeta| \rightarrow 1$,
that is, the
$\zeta$ representation is redefined {\em on\/} the unit circle
instead {\em inside}. The function $\Theta(f;\theta)$ of
Eq.\ (\ref{2.33})
then can be called the ``boundary function'' of the function
$f(\frac{1}{2},\zeta)$ analytic in the unit disk
[Eq.\ (\ref{2.16})]. The boundary function $\Theta(f;\theta)$
determines uniquely the analytic
representation $f(\frac{1}{2},\zeta)$ \cite{Vou93}:
	\begin{equation}
f(\mbox{{\small $\frac{1}{2}$}},\zeta)
=\sum_{n = 0}^{\infty} C_{n}(f) \zeta^{n} =
\int_{\theta_{0}}^{\theta_{0}+ 2\pi } d\theta \,
\frac{\Theta(f;\theta)}{1-\zeta e^{i\theta}} .
\label{2.35}
	\end{equation}
In general, the $\Theta(f;\theta)$ is given by Fourier series of the
form (\ref{2.33}). However, there are states for which this Fourier
series can be converted into a relatively
simple functional form. We call such states the {\em philophase
states\/}, and the phase-like states $|\frac{1}{2},\zeta\rangle$
are an example of them:
	\begin{equation}
\Theta(\mbox{{\small $\frac{1}{2}$}},\zeta;\theta) = \sqrt{2\pi}\,
\langle\theta|\mbox{{\small $\frac{1}{2}$}},\zeta\rangle
= \frac{\sqrt{1-|\zeta|^{2}}}{1-\zeta e^{-i\theta}} .
\label{2.36}
	\end{equation}

\subsection{Statistical and phase properties}
\label{cs_D}

The photon-number distribution of the CS $|k,\zeta\rangle$
is the negative binomial distribution \cite{Vou92,Agar}:
	\begin{equation}
P_{n}(k,\zeta) = |\langle n|k,\zeta\rangle|^{2}
= (1-|\zeta|^{2})^{2k}\frac{\Gamma(n+2k)}{n!\Gamma(2k)}
|\zeta|^{2n} .
\label{2.37}
	\end{equation}
The mean photon number $\langle\hat{n}\rangle{}_{k,\zeta}$ and the
intensity correlation function $g^{(2)}_{k,\zeta}$ are \cite{Vou92}
	\begin{equation}
\langle\hat{n}\rangle{}_{k,\zeta} =
2k\frac{|\zeta|^{2}}{1-|\zeta|^{2}} ,
\label{2.38}
	\end{equation}
	\begin{equation}
g^{(2)}_{k,\zeta} = \frac{ \langle\hat{n}^{2}\rangle{}_{k,\zeta}
- \langle\hat{n}\rangle{}_{k,\zeta} }{
\langle\hat{n}\rangle{}_{k,\zeta}^{2} }
= 1+ \frac{1}{2k} .
\label{2.39}
	\end{equation}
The photon-number distribution $P_{n}(k,\zeta)$ is super-Poissonian
$\left( g^{(2)}_{k,\zeta} > 1 \right)$, and for $k=\frac{1}{2}$ it
becomes
the thermal distribution \cite{Ahar,Vou92}.

Phase properties of a normalized photon state are
obtained by calculating expectation values of appropriate
{\em phase-related operators\/}. We define phase-related operators
as operators that can be written in the form of Fourier-like series
in the exponential phase operators \cite{LuPer,BBA_PR1},
	\begin{equation}
\hat{G} = \tilde{G}_{0}\hat{1} +
\sum_{n=1}^{\infty} \left[
\tilde{G}_{n}\left(\widehat{e^{-i\phi}}\right)^{n} +
\tilde{G}_{-n}\left(\widehat{e^{i\phi}}\right)^{n} \right] .
\label{2.40}
	\end{equation}
Fourier coefficients are given by
	\begin{equation}
 \tilde{G}_{n} = \frac{1}{2\pi}
\int_{\theta_{0}}^{\theta_{0}+ 2\pi } d\theta \,
G(\theta)\, e^{in\theta} ,
\label{2.41}
	\end{equation}
where $G(\theta)$ is a classical function corresponding to the
operator $\hat{G}$. It is easy to see that phase-related
operators have diagonal phase-state representation
\cite{LuPer,BBA_PR1},
	\begin{equation}
\hat{G} = \int_{\theta_{0}}^{\theta_{0}+ 2\pi }
d\theta \, G(\theta) |\theta\rangle\langle\theta| .
\label{2.42}
	\end{equation}
An example of such operators is the Susskind-Glogower cosine phase
operator $\hat{C}$ \cite{SG,CN} given by
	\begin{equation}
\hat{C} = \frac{1}{2} \left( \widehat{e^{i\phi}}
+\widehat{e^{-i\phi}} \right) =
\int_{\theta_{0}}^{\theta_{0}+ 2\pi } d\theta \, \!\cos
\theta |\theta\rangle\langle\theta| .
\label{2.43}
	\end{equation}
Functions $G(\theta)$ must be $2\pi$-periodic functions with
convergent Fourier series. For a function $G(\theta)$, which is not
intrinsically $2\pi$ periodic, we must use the periodic expansion
on the entire real axis, e.g.,
	\begin{equation}
\theta_{\rm per} =\theta\; \mbox{ for } \theta \in
[\theta_{0}, \theta_{0} + 2\pi )\;\;
\mbox{+ $2\pi$-periodic expansion on {\Bbb{R}}.}
 \label{2.44}
	\end{equation}
Then Hermitian phase operator $\hat{\phi}$ is given by
\cite{LuPer,BBA_PR1}
	\begin{equation}
\hat{\phi}=\int_{\theta_{0}}^{\theta_{0}+ 2\pi } d\theta \,
\theta_{\rm per} |\theta\rangle\langle\theta|
= \tilde{\theta}_{0} \hat{1} + \sum_{n=1}^{\infty} \left[
\tilde{\theta}_{n}\left(\widehat{e^{-i\phi}}\right)^{n} +
\tilde{\theta}_{-n}\left(\widehat{e^{i\phi}}\right)^{n} \right] ,
\label{2.45}
	\end{equation}
where
	\begin{equation}
\tilde{\theta}_{n} = \frac{1}{2\pi}
\int_{\theta_{0}}^{\theta_{0}+ 2\pi } d\theta \,
\theta \, e^{in\theta}
=\left\{ \begin{array}{ccl}
\theta_{0} + \pi ,               &   &  n=0 , \\
\frac{1}{in} e^{in\theta_{0}} ,  &   &  n \neq 0 .
   \end{array} \right.
\label{2.46}
	\end{equation}
The expectation value of a phase-related operator $\hat{G}$
over a photon state is
	\begin{equation}
\langle\hat{G}\rangle = {\rm Tr}\,
(\hat{\rho}\hat{G})
= \int_{\theta_{0}}^{\theta_{0}+ 2\pi } d\theta \,
G(\theta) \langle\theta|\hat{\rho}|\theta\rangle ,
\label{2.47}
	\end{equation}
where $\hat{\rho}$ is the density operator of the state,
and we have used the form (\ref{2.42}) for $\hat{G}$.

It is easy to see that such operators as $\hat{C}^{2}$,
$\hat{\phi}^{2}$, etc. cannot be written in the form
(\ref{2.40}) or
(\ref{2.42}). In order to calculate the expectation values of these
operators we use the {\em antinormal ordering\/} of the exponential
phase operators $\widehat{e^{i\phi}}$ and $\widehat{e^{-i\phi}}$,
defined as a procedure that places all raising operators
$\widehat{e^{-i\phi}}$ to the right of all  lowering
operators $\widehat{e^{i\phi}}$. The antinormal ordering restores
the unitarity for the exponential phase operators,
	\begin{equation}
\widehat{e^{i\phi}}\widehat{e^{-i\phi}} = {}^{\ast}_{\ast}
\widehat{e^{-i\phi}}\widehat{e^{i\phi}} {}^{\ast}_{\ast}
=\hat{1} ,
\label{2.48}
	\end{equation}
where two ${}^{\ast}_{\ast}$ on either side of an expression are
our notation of
the antinormal ordering. By applying the antinormal ordering to
the product of any two phase-related operators $\hat{G}$ and
$\hat{F}$, we obtain another phase-related operator
\cite{BBA_PR1}:
	\begin{equation}
{}^{\ast}_{\ast} \hat{G}\hat{F} \,{}^{\ast}_{\ast}
= \int_{\theta_{0}}^{\theta_{0}+ 2\pi }
d\theta \, G(\theta) F(\theta) |\theta\rangle\langle\theta| ,
\label{2.51}
	\end{equation}
where both $\hat{G}$ and $\hat{F}$ can be written in
the form (\ref{2.42}). Thus any two phase-related operators commute,
	\begin{equation}
{}^{\ast}_{\ast} [ \hat{G} , \hat{F} ]
{}^{\ast}_{\ast} = 0 ,
\label{2.49}
	\end{equation}
so, the phase variable is unique. The vacuum state may then be
described as a state of a random phase, similar to all other
number states. We have explained \cite{BBA_PR1} that by the
antinormal ordering we exclude nonrandom phase properties for the
vacuum. From Eq.\ (\ref{2.51}), one gets
	\begin{equation}
{}^{\ast}_{\ast} \hat{G}^{n} {}^{\ast}_{\ast} =
\int_{\theta_{0}}^{\theta_{0}+ 2\pi } d\theta \,
[G(\theta)]^{n} |\theta\rangle\langle\theta| .
\label{2.52}
	\end{equation}
We obtain, for example,
	\begin{equation}
{}^{\ast}_{\ast} \hat{C}^{2} {}^{\ast}_{\ast} =
\int_{\theta_{0}}^{\theta_{0}+ 2\pi } d\theta \, \cos^{2}\!\theta
|\theta\rangle\langle\theta| .
\label{2.53}
	\end{equation}

All the information about phase properties of a state is contained
in the phase distribution function $Q(\theta)$, given by
	\begin{equation}
Q(\theta) = \langle\theta|\hat{\rho}|\theta\rangle .
\label{2.54}
	\end{equation}
Then the expectation value of a phase-related operator
$\hat{G}$ is
	\begin{equation}
\langle\hat{G}\rangle =
\int_{\theta_{0}}^{\theta_{0}+ 2\pi } d\theta \,
G(\theta) Q(\theta) ,
\label{2.55}
	\end{equation}
and
	\begin{equation}
\langle{}^{\ast}_{\ast}\hat{G}\hat{F}\,
{}^{\ast}_{\ast}\rangle = \int_{\theta_{0}}^{\theta_{0}+ 2\pi }
d\theta \, G(\theta) F(\theta) Q(\theta) .
\label{2.56}
	\end{equation}
It has been shown recently \cite{BBA_PR1} that results for
phase-related expectation values obtained by using the antinormal
ordering are equivalent to those calculated in the frames of the
Pegg-Barnett formalism \cite{PB}
with proper limiting procedures \cite{VP}.
For a pure normalized state $|f\rangle$ of the form (\ref{2.15}),
the $Q(\theta)$ function is given by
	\begin{equation}
Q(f;\theta) = |\langle\theta|f\rangle|^{2} = \frac{1}{2\pi}
|\Theta(f;\theta)|^{2}
= \frac{1}{2\pi} \sum_{n,m=0}^{\infty} C_{n}(f) C_{m}^{\ast}(f)
e^{-i(n-m)\theta} .
\label{2.57}
	\end{equation}
Often we meet a special case, in which
	\begin{equation}
C_{n}(f) = |C_{n}(f)| \exp(in \bar{\varphi}_{f}) .
\label{2.58}
	\end{equation}
Then the phase distribution function has form
	\begin{equation}
Q(f;\theta) = \frac{1}{2\pi} \sum_{n=-\infty}^{\infty}
{\cal M}_{|n|}(f) \exp[-in(\theta-\bar{\varphi}_{f})]
= \frac{1}{2\pi} \left\{ 1+ 2 \sum_{n=1}^{\infty}
{\cal M}_{n}(f) \cos[n(\theta-\bar{\varphi}_{f})] \right\} ,
\label{2.59}
	\end{equation}
where
	\begin{equation}
{\cal M}_{n}(f) = \sum_{m = 0}^{\infty} |C_{m}(f) C_{m+n}(f)|,
\mbox{\hspace{1cm}}
 n \geq 0 .
\label{2.60}
	\end{equation}
Substituting into Eq.\ (\ref{2.55}) this expression for $Q(f;\theta)$
and Fourier series representing $G(\theta)$, we get
	\begin{equation}
\langle\hat{G}\rangle_{f} \equiv \langle
f|\hat{G}|f \rangle = \tilde{G}_{0}
+ \sum_{n=1}^{\infty} {\cal M}_{n}(f)
[ \tilde{G}_{n} \exp(-in\bar{\varphi}_{f}) + \tilde{G}_{-n}
\exp(in\bar{\varphi}_{f}) ] .
\label{2.61}
	\end{equation}
Fourier coefficients $\tilde{G}_{n}$ are given by Eq.\ (\ref{2.41}).
Expectation values are calculated for observables represented by
Hermitian operators. For a Hermitian phase-related operator
$\hat{G}$,
the function $G(\theta)$ is real, and $\tilde{G}_{-n} =
\tilde{G}_{n}^{\ast}$. Then
	\begin{equation}
\langle\hat{G}\rangle_{f} = \tilde{G}_{0} + 2
\sum_{n=1}^{\infty}
{\cal M}_{n}(f) {\rm Re}\, [ \tilde{G}_{n}
\exp(-in\bar{\varphi}_{f}) ] .
\label{2.62}
	\end{equation}
The expectation value of the Hermitian phase operator
$\hat{\phi}$ of Eq.\ (\ref{2.45}) is
	\begin{equation}
\langle\hat{\phi}\rangle_{f} = \theta_{0} + \pi + 2
\sum_{n=1}^{\infty}
\frac{1}{n} {\cal M}_{n}(f) \sin[n(\theta_{0}-\bar{\varphi}_{f})] .
\label{2.63}
	\end{equation}
The natural choice of the reference phase is
	\begin{equation}
\theta_{0} = \bar{\varphi}_{f} - \pi .
\label{2.64}
	\end{equation}
Then the expectation value of the phase observable is at the center
of the interval $[\theta_{0},\theta_{0}+2\pi)$:
	\begin{equation}
\langle\hat{\phi}\rangle_{f} = \bar{\varphi}_{f} =
\theta_{0} + \pi .
\label{2.65}
	\end{equation}
The antinormally ordered square of the Hermitian phase operator
is given by
	\begin{equation}
{}^{\ast}_{\ast} \hat{\phi}^{2} {}^{\ast}_{\ast} =
\int_{\theta_{0}}^{\theta_{0}+ 2\pi } d\theta \,
\theta^{2}_{\rm per} |\theta\rangle\langle\theta|
= \tilde{\theta}_{0}^{(2)} \hat{1} +
\sum_{n=1}^{\infty} \left[
\tilde{\theta}_{n}^{(2)} \left(\widehat{e^{-i\phi}}\right)^{n} +
\tilde{\theta}_{-n}^{(2)} \left(\widehat{e^{i\phi}}\right)^{n}
\right] ,
\label{2.66}
	\end{equation}
where
	\begin{equation}
\tilde{\theta}_{n}^{(2)} = \frac{1}{2\pi}
\int_{\theta_{0}}^{\theta_{0}+ 2\pi } d\theta \, \theta^{2} \,
e^{in\theta} =\left\{ \begin{array}{c}
\frac{4}{3}\pi^{2} +2\pi\theta_{0} + \theta_{0}^{2} ,
\;\;\;  n=0 , \\
 2e^{in\theta_{0}} \left( \frac{\pi +\theta_{0}}{in}
+ \frac{1}{n^{2}} \right) , \;\;\;  n \neq 0 .
   \end{array} \right.
\label{2.67}
	\end{equation}
With the choice (\ref{2.64}) the expectation value of
${}^{\ast}_{\ast} \hat{\phi}^{2} {}^{\ast}_{\ast}$ is
	\begin{equation}
\langle{}^{\ast}_{\ast}\hat{\phi}^{2}{}^{\ast}_{\ast}
\rangle_{f} = \frac{\pi^{2}}{3} + \bar{\varphi}_{f}^{2}
+ 4 \sum_{n=1}^{\infty} \frac{(-1)^{n}}{n^{2}} {\cal M}_{n}(f) ,
\label{2.68}
	\end{equation}
and the phase variance is
	\begin{equation}
{}^{\ast}_{\ast}(\Delta\phi)^{2}_{f}{}^{\ast}_{\ast} =
\langle{}^{\ast}_{\ast}\hat{\phi}^{2}{}^{\ast}_{\ast}
\rangle_{f}
- \langle\hat{\phi}\rangle_{f}^{2} = \frac{\pi^{2}}{3}
+ 4 \sum_{n=1}^{\infty} \frac{(-1)^{n}}{n^{2}} {\cal M}_{n}(f) .
\label{2.69}
	\end{equation}
In a similar way, we obtain
	\begin{equation}
\langle\hat{C}\rangle_{f} = {\cal M}_{1}(f) \cos
\bar{\varphi}_{f} ,
\label{2.70}
	\end{equation}
	\begin{equation}
\langle{}^{\ast}_{\ast}\hat{C}^{2}{}^{\ast}_{\ast}
\rangle_{f} = \frac{1}{2} [1-{\cal M}_{2}(f)]
+ {\cal M}_{2}(f) \cos^{2}\! \bar{\varphi}_{f} ,
\label{2.71}
	\end{equation}
	\begin{equation}
{}^{\ast}_{\ast}(\Delta C)^{2}_{f}{}^{\ast}_{\ast} =
\langle{}^{\ast}_{\ast}\hat{C}^{2}{}^{\ast}_{\ast}
\rangle_{f}
- \langle\hat{C}\rangle_{f}^{2} = \frac{1}{2}
[1-{\cal M}_{2}(f)]
+ [{\cal M}_{2}(f) - {\cal M}_{1}^{2}(f)] \cos^{2}\!
\bar{\varphi}_{f} .
\label{2.72}
	\end{equation}
When using the Susskind-Glogower sine phase operator
$\hat{S}$ \cite{SG,CN},
analogous formulas for $\langle\hat{S}\rangle_{f}$,
$\langle{}^{\ast}_{\ast}\hat{S}^{2}{}^{\ast}_{\ast}
\rangle_{f}$ and ${}^{\ast}_{\ast}(\Delta S)^{2}_{f}{}^{\ast}_{\ast}$
are obtained by replacing $\cos \bar{\varphi}_{f}$ into
$\sin \bar{\varphi}_{f}$ in Eqs. (\ref{2.70})-(\ref{2.72}).

Now we apply these general results to SU(1,1) CS $|k,\zeta\rangle$.
We get
	\begin{equation}
{\cal M}_{n}(k,\zeta) = |\zeta|^{n}
\frac{(1-|\zeta|^{2})^{2k}}{\Gamma(2k)}
\sum_{m = 0}^{\infty} \left[
\frac{\Gamma(m+n+2k)\Gamma(m+2k)}{(m+n)!m!}
\right]^{1/2} |\zeta|^{2m} .
\label{2.73}
	\end{equation}
The simplest case is $k=\frac{1}{2}$. Then
${\cal M}_{n}(\frac{1}{2},\zeta) = |\zeta|^{n}$, so,
	\begin{equation}
Q(\mbox{{\small $\frac{1}{2}$}},\zeta;\theta) = \frac{1}{2\pi}
\frac{1-|\zeta|^{2}}{1+|\zeta|^{2}
-2|\zeta| \cos(\theta-\bar{\varphi}_{\zeta}) } .
\label{2.74}
	\end{equation}
Here
	\begin{equation}
\langle\hat{\phi}\rangle_{\zeta} = \bar{\varphi}_{\zeta}
= {\rm arg}\,\zeta
\label{2.75}
	\end{equation}
for all values of $k$. Also, we obtain
	\begin{equation}
{}^{\ast}_{\ast}(\Delta\phi)^{2}_{\frac{1}{2},\zeta}{}^{\ast}_{\ast}
= \frac{\pi^{2}}{3}
+ 4 \sum_{n=1}^{\infty} \frac{(-1)^{n}}{n^{2}} |\zeta|^{n} ,
\label{2.76}
	\end{equation}
	\begin{equation}
\langle\hat{C}\rangle_{\frac{1}{2},\zeta} =
|\zeta| \cos \bar{\varphi}_{\zeta} ,
\label{2.77}
	\end{equation}
	\begin{equation}
{}^{\ast}_{\ast}(\Delta C)^{2}_{\frac{1}{2},\zeta}{}^{\ast}_{\ast} =
\frac{1}{2} (1-|\zeta|^{2}) .
\label{2.78}
	\end{equation}
The mean photon number $\langle\hat{n}\rangle_{\frac{1}{2},\zeta} =
|\zeta|^{2} (1-|\zeta|^{2})^{-1}$
shows that the quantum limit (small excitations) is obtained
for $|\zeta| \ll 1$ and the classical limit (large excitations)
corresponds to $|\zeta| \rightarrow 1$.
In the quantum limit, when $|\zeta| \rightarrow 0$,
${}^{\ast}_{\ast}(\Delta\phi)^{2}_{\frac{1}{2},\zeta}
{}^{\ast}_{\ast} \approx \pi^{2}/3$
and ${}^{\ast}_{\ast}(\Delta C)^{2}_{\frac{1}{2},\zeta}
{}^{\ast}_{\ast} \approx \frac{1}{2}$, that corresponds
to the uniform phase distribution, i.e.,
$Q(\mbox{{\small $\frac{1}{2}$}},\zeta;\theta) \approx
(2\pi)^{-1}$. In the classical limit, when $|\zeta| \rightarrow 1$,
both ${}^{\ast}_{\ast}(\Delta\phi)^{2}_{\frac{1}{2},\zeta}
{}^{\ast}_{\ast}$ and
${}^{\ast}_{\ast}(\Delta C)^{2}_{\frac{1}{2},\zeta}{}^{\ast}_{\ast}$
tend to zero, that corresponds
to a state with a perfectly defined phase. In this limit
the $Q(\mbox{{\small $\frac{1}{2}$}},\zeta;\theta)$ function is
very narrow.
By using the expression (\ref{2.73})
for ${\cal M}_{n}(k,\zeta)$, we can calculate numerically the phase
distribution function $Q(k,\zeta;\theta)$, given generally by Eq.\
(\ref{2.59}). The larger
values of $k$ are, the narrower the $Q(k,\zeta;\theta)$ function is,
and the better the phase of the state is defined, for given
$|\zeta|$.
We can calculate numerically also the
${}^{\ast}_{\ast}(\Delta\phi)^{2}_{k,\zeta}{}^{\ast}_{\ast}$
and ${}^{\ast}_{\ast}(\Delta C)^{2}_{k,\zeta}{}^{\ast}_{\ast}$,
by using Eqs.\ (\ref{2.69}) and
(\ref{2.72}), respectively, for different values of $k$.
The phase variance ${}^{\ast}_{\ast}(\Delta\phi)^{2}_{k,\zeta}
{}^{\ast}_{\ast}$ is
independent of $\bar{\varphi}_{\zeta} = {\rm arg}\,\zeta$.
When $|\zeta| \rightarrow 0$, the phase variance tends to the random
value $\pi^{2}/3$, and when $|\zeta| \rightarrow 1$, it tends
to zero, for all values of $k$. The larger values of $k$ are, the
smaller the phase variance is, for given $|\zeta|$.
A similar situation is with the cosine
variance ${}^{\ast}_{\ast}(\Delta C)^{2}_{k,\zeta}{}^{\ast}_{\ast}$.
When $|\zeta| \rightarrow 0$, the
cosine variance
tends to the random value $\frac{1}{2}$, and when
$|\zeta| \rightarrow 1$, it tends to zero.

\subsection{Number-phase uncertainty relations}
\label{cs_E}

The number and phase operators form a Heisenberg pair of canonically
conjugate observables \cite{GW}. The number-phase
commutation relation is given by \cite{LuPer,BBA_PR2}
	\begin{equation}
[\hat{n},\hat{\phi}] = i(1-2\pi|\theta_{0}
\rangle\langle\theta_{0}|) .
\label{2.79}
	\end{equation}
The additional term $-2\pi i|\theta_{0}\rangle\langle\theta_{0}|$
in this relation takes into account the periodicity of the phase.
Equation (\ref{2.79}) is a special case of the general commutation
relation between any phase-related operator $\hat{G}$ of the form
(\ref{2.42}) and the number operator $\hat{n}$ \cite{LuPer,BBA_PR2}:
	\begin{equation}
[\hat{n} , \hat{G}] =
i \int_{\theta_{0}}^{\theta_{0}+ 2\pi } d\theta \,
|\theta\rangle\langle\theta| \frac{d}{d\theta} G(\theta) .
\label{crel}
	\end{equation}
For the phase operator $\hat{\phi}$ of Eq.\ (\ref{2.45}),
we get
	\begin{equation}
\frac{d}{d\theta} \theta_{\rm per} =
1-2\pi \delta(\theta-\theta_{0}) ,
	\end{equation}
and the relation (\ref{2.79}) follows immediately from Eq.\
(\ref{crel}).
The number-phase uncertainty relation reads
	\begin{equation}
(\Delta n)^{2} {}^{\ast}_{\ast}(\Delta\phi)^{2}{}^{\ast}_{\ast}
\geq \frac{1}{4}
[1-2\pi Q(\theta_{0})]^{2} .
\label{2.80}
	\end{equation}
We define the following function:
	\begin{equation}
{\cal V} \equiv \frac{(\Delta n)^{2}
{}^{\ast}_{\ast}(\Delta\phi)^{2}{}^{\ast}_{\ast}}{[1-2\pi
Q(\theta_{0})]^{2}} \geq \frac{1}{4} .
\label{2.81}
	\end{equation}
By evaluating this function, we can investigate the number-phase
uncertainty relation (\ref{2.80}) for various photon states.

The number variance for the SU(1,1) CS can be calculated from the
function $P_{n}(k,\zeta)$ of Eq.\ (\ref{2.37}). One gets
	\begin{equation}
(\Delta n)^{2}_{k,\zeta} = 2k\frac{|\zeta|^{2}}{(1-|\zeta|^{2})^{2}}.
\label{2.82}
	\end{equation}
By using the standard choice (\ref{2.64}) for $\theta_{0}$, we obtain
from Eq.\ (\ref{2.59})
	\begin{equation}
Q(k,\zeta;\theta_{0}) = \frac{1}{2\pi} \left[ 1+2\sum_{n=1}^{\infty}
(-1)^{n} {\cal M}_{n}(k,\zeta) \right] ,
\label{2.83}
	\end{equation}
where coefficients ${\cal M}_{n}(k,\zeta)$ are given by Eq.\
(\ref{2.73}). The phase variance can be calculated using Eq.\
(\ref{2.69}). The simplest case is $k=\frac{1}{2}$. Then
${}^{\ast}_{\ast}(\Delta\phi)^{2}_{\frac{1}{2},\zeta}
{}^{\ast}_{\ast}$ is given by Eq.\ (\ref{2.76}),
and
	\begin{equation}
Q(\mbox{{\small $\frac{1}{2}$}},\zeta;\theta_{0}) = \frac{1}{2\pi}
\frac{1-|\zeta|}{1+|\zeta|} .
\label{2.84}
	\end{equation}
Then Eq.\ (\ref{2.81}) reads
	\begin{equation}
{\cal V}(\mbox{{\small $\frac{1}{2}$}},\zeta) =
\frac{1}{4(1-|\zeta|)^{2}} \left[
\frac{\pi^{2}}{3} + 4\sum_{n=1}^{\infty} \frac{(-1)^{n}}{n^{2}}
|\zeta|^{n} \right] .
\label{2.85}
	\end{equation}
For $|\zeta| \rightarrow 0$, this function tends to the
random-phase value
$\pi^{2}/12$, while for $|\zeta|$ close to 1, it blows up. Numerical
calculations show that this limiting behavior is universal for all
values of $k$. We see that, for small
values of $k$, the CS $| k,\zeta \rangle$ are far from satisfying an
equality in
the number-phase uncertainty relation (\ref{2.80}). However, for a
large $k$, the function ${\cal V}(k,\zeta)$ is, in an intermediate
range of $|\zeta|$ values, close to its limit $\frac{1}{4}$.
The fact that the ${\cal V}(k,\zeta)$ function blows up for
$|\zeta| \rightarrow 1$ is, at first sight, somewhat strange and
contrary to the standard conception of ``the classical limit''
where uncertainties can be neglected. Indeed, the phase of the
SU(1,1) CS becomes perfectly defined in the limit
$|\zeta| \rightarrow 1$.
But the situation with the photon statistics is absolutely
different. The relative photon-number uncertainty for the
$|k,\zeta\rangle$ states is
	\begin{equation}
\frac{\Delta n_{k,\zeta}}{\langle\hat{n}\rangle_{k,\zeta}}
= \frac{1}{\sqrt{2k}\,|\zeta|} .
	\end{equation}
For $|\zeta| \rightarrow 1$, we obtain nothing similar to the
classical zero uncertainty (unless $k \rightarrow \infty$).
For example, for $k=\frac{1}{2}$, the relative photon-number
uncertainty
tends to unity in the limit $|\zeta| \rightarrow 1$. This follows
from the fact that the $|\frac{1}{2},\zeta\rangle$ states have
thermal photon-number distribution, though they are pure states.

One can also discuss the uncertainty relations for the number and
cosine or sine phase operators. From the commutation relations
	\begin{equation}
[\hat{n},\hat{C}] = -i\hat{S} ,
\mbox{\hspace{1.2cm}}
[\hat{n},\hat{S}] = i\hat{C} ,
\label{2.86}
	\end{equation}
one can deduce the uncertainty relations
	\begin{mathletters}
\label{2.87}
	\begin{equation}
(\Delta n)^{2} {}^{\ast}_{\ast}(\Delta C)^{2}{}^{\ast}_{\ast}
\geq \frac{1}{4}
\langle \hat{S} \rangle^{2} ,
\label{2.87a}
	\end{equation}
	\begin{equation}
(\Delta n)^{2} {}^{\ast}_{\ast}(\Delta S)^{2}{}^{\ast}_{\ast}
\geq\frac{1}{4}
\langle \hat{C} \rangle^{2} .
\label{2.87b}
	\end{equation}
	\end{mathletters}
In order to investigate these uncertainty relations, it is convenient
to define the functions
	\begin{mathletters}
\label{2.88}
	\begin{equation}
{\cal R}_{1} \equiv \frac{(\Delta n)^{2}
{}^{\ast}_{\ast}(\Delta C)^{2}{}^{\ast}_{\ast}}{\langle
\hat{S} \rangle^{2}} \geq \frac{1}{4} ,
\label{2.88a}
	\end{equation}
	\begin{equation}
{\cal R}_{2} \equiv \frac{(\Delta n)^{2}
{}^{\ast}_{\ast}(\Delta S)^{2}{}^{\ast}_{\ast}}{\langle
\hat{C} \rangle^{2}} \geq \frac{1}{4} .
\label{2.88b}
	\end{equation}
	\end{mathletters}
For a pure state $|f\rangle$, satisfying the condition (\ref{2.58}),
we get
	\begin{equation}
{\cal R}_{1}(f) = (\Delta n)^{2}_{f} \frac{\frac{1}{2}
[1-{\cal M}_{2}(f)]
+ [ {\cal M}_{2}(f) - {\cal M}_{1}^{2}(f)]
\cos^{2}\!\bar{\varphi}_{f}
}{ {\cal M}_{1}^{2}(f)\sin^{2}\!\bar{\varphi}_{f} } ,
\label{2.89}
	\end{equation}
and the ${\cal R}_{2}(f)$ is given by interchange of
$\cos \bar{\varphi}_{f}$ and $\sin \bar{\varphi}_{f}$.
For $\bar{\varphi}_{f} = \pi/4$, the functions
${\cal R}_{1}(f)$ and ${\cal R}_{2}(f)$ coincide and come to the
function
	\begin{equation}
{\cal U}(f) \equiv (\Delta n)^{2}_{f} \frac{
{}^{\ast}_{\ast}(\Delta C)^{2}_{f}{}^{\ast}_{\ast} +
{}^{\ast}_{\ast}(\Delta S)^{2}_{f}{}^{\ast}_{\ast}
}{ \langle \hat{C} \rangle^{2}_{f} + \langle \hat{S}
\rangle^{2}_{f} }
= (\Delta n)^{2}_{f} \frac{ 1-{\cal M}_{1}^{2}(f)
}{ {\cal M}_{1}^{2}(f) } .
\label{2.90}
	\end{equation}

We return now to the CS $| k,\zeta \rangle$. When $k=\frac{1}{2}$,
we get
	\begin{equation}
{\cal R}_{1}(\mbox{{\small $\frac{1}{2}$}},\zeta) =
[ 2(1-|\zeta|^{2}) \sin^{2}\!\bar{\varphi}_{\zeta} ]^{-1} .
\label{2.91}
	\end{equation}
For $|\zeta| \rightarrow 0$, this function tends to the value
$[ 2\sin^{2}\!\bar{\varphi}_{\zeta} ]^{-1} \geq \frac{1}{2}$, while
for $|\zeta|$ close
to 1, it blows up. The SU(1,1) CS with other values of $k$ behave
similarly. When $k$ is small, the CS $| k,\zeta \rangle$ give
a strong inequality in both
uncertainty relations (\ref{2.87a}) and (\ref{2.87b}). However, for
a large $k$, the function ${\cal U}(k,\zeta)$ is, in an intermediate
range of $|\zeta|$ values, close to its limit $\frac{1}{4}$.
It means that the
SU(1,1) CS with large values of $k$ approach to satisfy, for some
$|\zeta|$ values, an equality in the number-cosine and number-sine
uncertainty relations. The functions ${\cal V}(k,\zeta)$ and
${\cal U}(k,\zeta)$
behave very similarly. Therefore, the number-phase uncertainty
relations can be studied using the Hermitian phase operator
$\hat{\phi}$ as well as using the cosine or sine phase operators.

\section{The Barut-Girardello states}
\label{sec:bg}

In this section we use results of Barut and Girardello (BG), who have
constructed \cite{BG} the eigenstates of the lowering generator
$\hat{K}_{-}(k)$,
	\begin{equation}
\hat{K}_{-}(k) |k,z\rangle = z |k,z\rangle ,
\label{3.1}
	\end{equation}
where $z$ is an arbitrary complex number. The normalized BG states
can be decomposed over the number-state basis,
	\begin{equation}
|k,z\rangle = \frac{z^{k-1/2}}{\sqrt{I_{2k-1}(2|z|)}}
\sum_{n = 0}^{\infty} \frac{z^{n}}{\sqrt{n!\Gamma(n+2k)}} |n\rangle ,
\label{3.2}
	\end{equation}
where $I_{\nu}$ is the $\nu$-order modified Bessel function of the
first kind. The BG states are not orthogonal to each other,
	\begin{equation}
\langle k,z_{1}|k,z_{2} \rangle = \frac{ I_{2k-1}
(2\sqrt{z_{1}^{\ast}z_{2}})
}{ \sqrt{ I_{2k-1}(2|z_{1}|) I_{2k-1}(2|z_{2}|) } } .
\label{3.3}
	\end{equation}
The BG states have a simple representation in the SU(1,1)
coherent-state basis:
	\begin{equation}
{\cal Z}(k,z,\zeta) = (1-|\zeta|^{2})^{-k}
\langle k,\zeta^{\ast} | k,z\rangle
= \frac{z^{k-1/2}}{\sqrt{ I_{2k-1}(2|z|) \Gamma(2k) }}
\exp(z\zeta)  .
	\end{equation}
In the following discussion we will consider various properties of
these states.

\subsection{The identity resolution and the analytic representation
on the complex plane}
\label{bg_A}

In order to prove that the BG states resolve the identity, one must
find the measure $d\mu(k,z)$ such that
	\begin{equation}
\int d\mu(k,z) |k,z \rangle\langle k,z| = \hat{1} .
\label{3.4}
	\end{equation}
Writing $d\mu(k,z) = \mu(k,|z|) d^{2}\! z$ and integrating over the
whole $z$ plane, we find
	\begin{equation}
2\pi \sum_{n = 0}^{\infty} \frac{ |n \rangle\langle n|
}{ n!\Gamma(n+2k) } \int_{0}^{\infty}
d|z|\, \mu(k,|z|) \frac{ |z|^{2n+2k} }{ I_{2k-1}(2|z|) } = \hat{1} .
\label{3.5}
	\end{equation}
By using the following formula \cite{AS}
	\begin{equation}
\int_{0}^{\infty} dt\, t^{a} K_{b}(t) = 2^{a-1} \Gamma \left(
\frac{a+b+1}{2} \right) \Gamma \left( \frac{a-b+1}{2} \right) ,
\mbox{\hspace{1cm}} {\rm Re}\,(a\pm b) > -1 ,
\label{3.6}
	\end{equation}
where $K_{\nu}$ is the $\nu$-order modified Bessel function of the
second kind, we see that Eqs.\ (\ref{3.4}) and (\ref{3.5}) are valid
for
	\begin{equation}
d\mu(k,z) = \frac{2}{\pi} K_{2k-1}(2|z|) I_{2k-1}(2|z|) d^{2}\! z .
\label{3.7}
	\end{equation}
Thus the BG states form, for each allowed value of $k$,
an overcomplete basis in the harmonic oscillator Hilbert space.

The harmonic oscillator Hilbert space can be represented as the
Hilbert space of entire functions $f(k,z)$, which are analytic over
the whole $z$ plane. For a normalized state $|f\rangle$ of the form
(\ref{2.15}), we get
	\begin{equation}
f(k,z) = \frac{\sqrt{I_{2k-1}(2|z|)}}{z^{k-1/2}}
\langle k,z^{\ast}|f\rangle = \sum_{n = 0}^{\infty}
\frac{C_{n}(f)
}{ \sqrt{n!\Gamma(n+2k)} } z^{n} ,
\label{3.9}
	\end{equation}
and this state can be represented in the BG basis:
	\begin{equation}
|f\rangle = \int d\mu(k,z) \frac{(z^{\ast})^{k-1/2}
}{\sqrt{I_{2k-1}(2|z|)}} f(k,z^{\ast}) |k,z\rangle .
\label{3.10}
	\end{equation}
The SU(1,1) generators $\hat{K}_{\pm}(k)$ and $\hat{K}_{3}(k)$
act on the Hilbert space of entire
functions $f(k,z)$ as linear operators \cite{BG}:
	\begin{equation}
\hat{K}_{+}(k) =  z ,  \;\;\;\;\;
 \hat{K}_{-}(k) = 2k \frac{d}{dz} + z \frac{d^{2}}{dz^{2}} ,
\;\;\;\;\;
 \hat{K}_{3}(k)  = z \frac{d}{dz} + k  .
\label{3.11}
	\end{equation}

\subsection{Statistical and phase properties and the number-phase
uncertainty relations}
\label{bg_D}

The photon-number distribution of the BG states $|k,z\rangle$,
	\begin{equation}
 P_{n}(k,z) = |\langle n|k,z \rangle|^{2} = \frac{ |z|^{2n+2k-1}
}{ I_{2k-1}(2|z|) n! \Gamma(n+2k) } ,
\label{3.12}
	\end{equation}
is sub-Poissonian \cite{BBA_QO}. The expectation values for
the number operator and its square can be easily calculated,
	\begin{eqnarray}
\langle\hat{n}\rangle_{k,z} & = & |z|
\frac{ I_{2k}(2|z|) }{ I_{2k-1}(2|z|) } ,
\label{3.13}  \\
 \langle\hat{n}^{2}\rangle_{k,z} & = & |z|^{2} - (2k-1)|z|
\frac{ I_{2k}(2|z|) }{ I_{2k-1}(2|z|) } .
\label{3.14}
	\end{eqnarray}
The intensity correlation function is given by
	\begin{equation}
g^{(2)}_{k,z} = \frac{ \langle\hat{n}^{2}\rangle_{k,z} -
\langle\hat{n}\rangle_{k,z}
}{ \langle\hat{n}\rangle_{k,z}^{2} }
= \frac{ I_{2k-1}(2|z|) }{ I_{2k}(2|z|) }  \left[
\frac{ I_{2k-1}(2|z|) }{ I_{2k}(2|z|) } - \frac{2k}{|z|} \right] .
\label{3.15}
	\end{equation}
In the quantum limit $|z| \ll 1$, we get approximately
	\begin{equation}
g^{(2)}_{k,z} \approx \frac{2k}{2k+1} .
\label{3.18}
	\end{equation}
We see that the maximal antibunching $g^{(2)}_{k,z} =\frac{1}{2}$
is achieved for $k=\frac{1}{2}$, while for large values of $k$, the
$g^{(2)}_{k,z}$ approaches unity. Another interesting feature is
that the intensity correlation function $g^{(2)}_{k,z}$
of Eq.\ (\ref{3.18}) is equal to the reciprocal of the coherent-state
intensity correlation function $g^{(2)}_{k,\zeta}$,
given by Eq.\ (\ref{2.39}),
	\begin{equation}
g^{(2)}_{k,z} \approx \left[ g^{(2)}_{k,\zeta} \right]^{-1} ,
\mbox{\hspace{1cm}} |z| \ll 1 .
\label{3.19}
	\end{equation}
It can be verified that the $g^{(2)}_{k,z}$ is always
less than unity, so the BG states have sub-Poissonian photon
statistics. Therefore, we call these states (which are,
in many aspects, similar to the Glauber coherent states) the
subcoherent states. The detail discussion of statistical
properties of the BG states is given in Ref.\ \cite{BBA_QO}.
Also, it is shown there that the BG basis can be used to construct
a diagonal representation of the density operator (the so-called
subcoherent $P$-representation), which is shown to be well-behaved
for nonclassical photon states.

According to the general results of Sec. \ref{cs_D}, we investigate
here phase properties of the BG subcoherent states. By using
Eq.\ (\ref{2.60}), we find
	\begin{equation}
{\cal M}_{n}(k,z) = \frac{ |z|^{n+2k-1} }{ I_{2k-1}(2|z|) }
\sum_{m = 0}^{\infty}
\frac{ |z|^{2m} }{ [m!(m+n)!\Gamma(m+2k)\Gamma(m+n+2k)]^{1/2} } .
\label{3.39}
	\end{equation}
Then the phase distribution function $Q(k,z;\theta)$ can be
calculated
from Eq.\ (\ref{2.59}). In the simplest case $k=\frac{1}{2}$, we get
	\begin{equation}
{\cal M}_{n}(\mbox{{\small $\frac{1}{2}$}},z) =
\frac{ I_{n}(2|z|) }{ I_{0}(2|z|) } ,
\label{3.40}
	\end{equation}
and
	\begin{equation}
Q(\mbox{{\small $\frac{1}{2}$}},z;\theta) =
\frac{1}{2\pi I_{0}(2|z|)} \sum_{n=-\infty}^{\infty}
I_{|n|}(2|z|) \exp [-in(\theta-\bar{\varphi}_{z})] .
\label{3.41}
	\end{equation}
Here $\langle\hat{\phi}\rangle_{z} = \bar{\varphi}_{z}
= {\rm arg}\, z$ for all values
of $k$. By using the expansion \cite{AS}
	\begin{equation}
e^{x\cos \theta} = \sum_{n=-\infty}^{\infty} I_{|n|}(x)
e^{-in\theta} ,
\label{3.42}
	\end{equation}
we can write
	\begin{equation}
Q(\mbox{{\small $\frac{1}{2}$}},z;\theta) =
\frac{ \exp [2|z|\cos(\theta-\bar{\varphi}_{z})]
}{ 2\pi I_{0}(2|z|) }  .
\label{3.43}
	\end{equation}
This result can be also obtained by noting that
	\begin{equation}
\Theta(\mbox{{\small $\frac{1}{2}$}},z;\theta) =
\frac{1}{\sqrt{I_{0}(2|z|)}}
\sum_{n = 0}^{\infty} \frac{z^{n}}{n!} e^{-in\theta}
=  \frac{\exp \left( z e^{-i\theta} \right) }{\sqrt{I_{0}(2|z|)}} .
\label{3.44}
	\end{equation}
We see that the BG subcoherent states $|\frac{1}{2},z\rangle$ are an
example of
philophase states. In the next section we will show that the states
$|\frac{1}{2},z\rangle$ are a special case of a wide class of
philophase states,
all of them have the phase distribution function of the form
(\ref{3.43}). This phase distribution function is the same as the
classical von Mises distribution \cite{Rao}.
In the quantum limit  $|z| \ll 1$, the
$Q(\mbox{{\small $\frac{1}{2}$}},z;\theta)$ tends to the
uniform phase distribution, while in the classical limit $|z| \gg 1$,
it is narrow. Also, we calculate
	\begin{equation}
{}^{\ast}_{\ast}(\Delta\phi)^{2}_{\frac{1}{2},z}{}^{\ast}_{\ast} =
\frac{\pi^{2}}{3}
+ \frac{4}{I_{0}(2|z|)} \sum_{n=1}^{\infty} \frac{(-1)^{n}}{n^{2}}
I_{n}(2|z|) ,
\label{3.45}
	\end{equation}
	\begin{equation}
\langle\hat{C}\rangle_{\frac{1}{2},z} =
\frac{ I_{1}(2|z|) }{ I_{0}(2|z|) }
\cos \bar{\varphi}_{z} ,
\label{3.46}
	\end{equation}
	\begin{equation}
{}^{\ast}_{\ast}(\Delta C)^{2}_{\frac{1}{2},z}{}^{\ast}_{\ast} =
\frac{1}{2} \left[ 1 -
\frac{ I_{2}(2|z|) }{ I_{0}(2|z|) } \right] + \left[
\frac{ I_{2}(2|z|) }{ I_{0}(2|z|) } -
\frac{ I_{1}^{2}(2|z|) }{ I_{0}^{2}(2|z|) } \right]
\cos^{2}\! \bar{\varphi}_{z} .
\label{3.47}
	\end{equation}
In the quantum limit  $|z| \ll 1$, we get approximately
	\begin{eqnarray}
{}^{\ast}_{\ast}(\Delta\phi)^{2}_{\frac{1}{2},z}{}^{\ast}_{\ast} &
\approx & \frac{\pi^{2}}{3} -4|z| ,
\label{3.48} \\
{}^{\ast}_{\ast}(\Delta C)^{2}_{\frac{1}{2},z}{}^{\ast}_{\ast} &
\approx & \frac{1}{2} \left( 1 -
\frac{|z|^{2}}{2} \right) - \frac{|z|^{2}}{2} \cos^{2}\!
\bar{\varphi}_{z} .
\label{3.49}
	\end{eqnarray}
Thus the phase variance
${}^{\ast}_{\ast}(\Delta\phi)^{2}_{\frac{1}{2},z}{}^{\ast}_{\ast}$
and the cosine variance
${}^{\ast}_{\ast}(\Delta C)^{2}_{\frac{1}{2},z}{}^{\ast}_{\ast}$
tend in the limit
$|z| \rightarrow 0$ to their random values $\pi^{2}/3$ and
$\frac{1}{2}$, respectively. In the classical limit of large $|z|$,
the ${}^{\ast}_{\ast}(\Delta\phi)^{2}_{\frac{1}{2},z}
{}^{\ast}_{\ast}$ tends to zero, and
the ${}^{\ast}_{\ast}(\Delta C)^{2}_{\frac{1}{2},z}
{}^{\ast}_{\ast}$ behaves similarly tending to the
value $(2|z|)^{-1} \sin^{2}\! \bar{\varphi}_{z}$. Thus,
for large values of
$|z|$, the phase of the  states $|\frac{1}{2},z\rangle$ is well
defined.
By using the expression (\ref{3.39}) for ${\cal M}_{n}(k,z)$, we can
calculate numerically the phase distribution function $Q(k,z,\theta)$
for different values of $k$. The larger values of $k$ are,
the flatter the $Q(k,z,\theta)$ is, and the worse the phase of the
state
is defined, for given $|z|$. This behavior is opposite to that of
the SU(1,1) CS. We calculate numerically also the variances
${}^{\ast}_{\ast}(\Delta\phi)^{2}_{k,z}{}^{\ast}_{\ast}$ and
${}^{\ast}_{\ast}(\Delta C)^{2}_{k,z}{}^{\ast}_{\ast}$
for different values of $k$. The phase variance
${}^{\ast}_{\ast}(\Delta\phi)^{2}_{k,z}{}^{\ast}_{\ast}$ is
independent of
$\bar{\varphi}_{z} = {\rm arg}\, z$. When $|z| \rightarrow 0$,
the phase variance tends to the random value $\pi^{2}/3$,
and when $|z|$ is large
($|z| \gg k$), it tends to zero, for all values of $k$. The larger
values of $k$ are, the larger the phase variance is, for given $|z|$.
We meet a similar
situation for the cosine variance ${}^{\ast}_{\ast}
(\Delta C)^{2}_{k,z}{}^{\ast}_{\ast}$. When $|z| \rightarrow 0$,
the cosine variance tends to the random value $\frac{1}{2}$, and
for large $|z|$ it tends to zero.

Now we discuss the number-phase uncertainty relations for the BG
subcoherent states. The number variance $(\Delta n)^{2}_{k,z}$ can
be deduced from Eqs.\ (\ref{3.13}) and (\ref{3.14}),
	\begin{equation}
(\Delta n)^{2}_{k,z} = |z|^{2} - (2k-1)|z| \frac{ I_{2k}(2|z|)
}{ I_{2k-1}(2|z|) } - |z|^{2} \frac{ I_{2k}^{2}(2|z|)
}{ I_{2k-1}^{2}(2|z|) } .
\label{3.50}
	\end{equation}
By using the standard choice (\ref{2.64}) for $\theta_{0}$, we obtain
from Eq.\ (\ref{2.59})
	\begin{equation}
Q(k,z;\theta_{0}) = \frac{1}{2\pi} \left[ 1 + 2\sum_{n=1}^{\infty}
(-1)^{n} {\cal M}_{n}(k,z) \right] ,
\label{3.51}
	\end{equation}
where the coefficients ${\cal M}_{n}(k,z)$ are given by Eq.\
(\ref{3.39}). The phase variance can be calculated using Eq.\
(\ref{2.69}). The simplest case is $k=\frac{1}{2}$. Then
${}^{\ast}_{\ast}(\Delta\phi)^{2}_{\frac{1}{2},z}{}^{\ast}_{\ast}$
is given by Eq. (\ref{3.45}), and
	\begin{equation}
Q(k,z;\theta_{0}) = \frac{ \exp(-2|z|) }{ 2\pi I_{0}(2|z|) } .
\label{3.52}
	\end{equation}
Thus we find the ${\cal V}$ function, defined by Eq.\ (\ref{2.81}),
	\begin{equation}
{\cal V}(\mbox{{\small $\frac{1}{2}$}},z) = |z|^{2} \left[ 1 -
\frac{ I_{1}^{2}(2|z|)
}{ I_{0}^{2}(2|z|) } \right] \left[ 1 - \frac{ \exp(-2|z|)
}{ 2\pi I_{0}(2|z|) } \right]^{-2} \left[ \frac{\pi^{2}}{3} +
\frac{4}{I_{0}(2|z|)} \sum_{n=1}^{\infty} \frac{(-1)^{n}}{n^{2}}
I_{n}(2|z|) \right] .
\label{3.53}
	\end{equation}
In the quantum limit $|z| \ll 1$, the function ${\cal V}
(\mbox{{\small $\frac{1}{2}$}},z)$ is given approximately by
	\begin{equation}
{\cal V}(\mbox{{\small $\frac{1}{2}$}},z)
\approx \frac{\pi^{2}}{12} - |z| \left( 1 -
\frac{\pi^{2}}{12} \right) .
\label{3.54}
	\end{equation}
In the classical limit $|z| \gg 1$, the ${\cal V}(\mbox{{\small
$\frac{1}{2}$}},z)$ tends to
its minimal possible value $\frac{1}{4}$. Numerical calculations
show that
this behaviour is universal for all values of $k$.
The larger values of $k$ are, the slower the function
${\cal V}(k,z)$ approaches, as $|z|$ increases, its limit
$\frac{1}{4}$.

An analogous situation is with the number-cosine and number-sine
uncertainty relations. For $k=\frac{1}{2}$, we get
	\begin{equation}
{\cal R}_{1}(\mbox{{\small $\frac{1}{2}$}},z) =
\frac{|z|^{2}}{\sin^{2}\!\bar{\varphi}_{z}}
\left[ \frac{I_{0}^{2}(2|z|)}{I_{1}^{2}(2|z|)} - 1 \right]
\left\{ \frac{1}{2} \left[ 1 - \frac{I_{2}(2|z|)}{I_{0}(2|z|)}
\right]
+ \cos^{2}\!\bar{\varphi}_{z} \left[ \frac{I_{2}(2|z|)}{I_{0}(2|z|)}
- \frac{I_{1}^{2}(2|z|)}{I_{0}^{2}(2|z|)} \right] \right\} ,
\label{3.55}
	\end{equation}
and the ${\cal R}_{2}(\frac{1}{2},z)$ is obtained by interchanging
$\cos \bar{\varphi}_{z}$ and $\sin \bar{\varphi}_{z}$.
In the quantum limit $|z| \ll 1$, we find
	\begin{equation}
{\cal R}_{1}(\mbox{{\small $\frac{1}{2}$}},z)
\approx (2\sin^{2}\!\bar{\varphi}_{z})^{-1} \geq \frac{1}{2} ,
\label{3.56}
	\end{equation}
while in the classical limit $|z| \gg 1$, the result is
	\begin{equation}
{\cal R}_{1}(\mbox{{\small $\frac{1}{2}$}},z) \approx \frac{1}{4}
\left( 1 + \frac{1}{2|z|} \right) .
\label{3.57}
	\end{equation}
We see that, for large $|z|$, the ${\cal R}_{1}(\frac{1}{2},z)$
approaches
its minimal possible value $\frac{1}{4}$. For other values of $k$
we meet
a similar behaviour. The larger
values of $k$ are, the slower an equality is achieved, as $|z|$
increases, in the number-cosine and number-sine uncertainty relations.
Generally, the number-phase uncertainty relations for the BG
subcoherent states behave very differently from those ones for the
SU(1,1) CS. However, we again see that the functions ${\cal V}$ and
${\cal U}$ behave very similarly. Hence, it is not very important
what is the phase function that we choose for investigating
number-phase uncertainty relations. This fact is a direct result
of the unique phase definition in the antinormal ordering formalism.
We can conclude by noting that the subcoherent states
$|k,z\rangle$ (especially $|\frac{1}{2},z\rangle$) are, for
$|z| \gg k$, a good
approximation to the number-phase intelligent states, i.e., they
tend to satisfy an equality in the number-phase uncertainty relation.
This property remains valid also regarding to the number-cosine
and number-sine uncertainty relations.

\section{Philophase states}
\label{sec:pp}

In the preceding sections we have seen
that the states $|\frac{1}{2},\zeta\rangle$ and
$|\frac{1}{2},z\rangle$ are two examples of philophase
states, i.e., states for which the phase-state representation
function $\Theta(\theta)$, defined by Eq.\ (\ref{2.33}) as Fourier
series, can be converted into a relatively simple functional form.
The function $\Theta(\frac{1}{2},\zeta;\theta)$ of the SU(1,1) CS
$|\frac{1}{2},\zeta\rangle$ is
given by Eq.\ (\ref{2.36}), and the function
$\Theta(\frac{1}{2},z;\theta)$ of
the BG subcoherent states $|\frac{1}{2},z\rangle$ is given
by Eq.\ (\ref{3.44}).
The philophase states $|\frac{1}{2},\zeta\rangle$ [phase-like
SU(1,1) CS] are the eigenstates of the
exponential phase operator $\widehat{e^{i\phi}}$
[see Eq.\ (\ref{2.28})],
and the philophase states $|\frac{1}{2},z\rangle$
(BG subcoherent states) are, by the definition
(\ref{3.1}), the eigenstates of the lowering generator
$\hat{K}_{-}(\frac{1}{2}) = \widehat{e^{i\phi}}\hat{n}$.
We can consider a generalized operator, defined by
	\begin{equation}
\hat{Z}(\sigma) \equiv \hat{K}_{-}(\mbox{{\small
$\frac{1}{2}$}}) + \sigma \widehat{e^{i\phi}} = \widehat{e^{i\phi}}
(\hat{n} + \sigma) ,
\label{4.1}
	\end{equation}
where $\sigma$ is, generally, any complex number. The eigenstates
of this operator are obtained from the following eigenvalue
equation:
	\begin{equation}
\hat{Z}(\sigma) |z,\sigma\rangle = z |z,\sigma\rangle ,
\label{4.2}
	\end{equation}
where $z$ is an arbitrary complex number. The states
$|z,\sigma\rangle$ can be decomposed over the number-state basis:
	\begin{equation}
|z,\sigma\rangle = \sum_{n = 0}^{\infty} C_{n}(z,\sigma) |n\rangle .
\label{4.3}
	\end{equation}
{}From the eigenvalue equation (\ref{4.2}) we deduce the following
recursion relation for the $C_{n}(z,\sigma)$:
	\begin{equation}
C_{n+1}(z,\sigma) = \frac{z}{n+\sigma+1} C_{n}(z,\sigma) ,
\mbox{\hspace{1cm}} n \geq 0 .
\label{4.4}
	\end{equation}
The solution of this equation is
	\begin{equation}
C_{n}(z,\sigma) = C_{0}(z,\sigma) \frac{z^{n}}{\Gamma(n+\sigma+1)} ,
\label{4.5}
	\end{equation}
where $C_{0}(z,\sigma)$ remains to be determined from the
normalization
condition. In the following discussion we restrict ourselves to the
relatively simple but important case of integer $\sigma$.
When $\sigma$ is zero, we obtain the BG subcoherent states
$|\frac{1}{2},z\rangle$,
whose properties have been discussed in detail in the preceding
section. It is convenient to distinguish between positive and
negative values of $\sigma$, since properties of the
$|z,\sigma\rangle$ states are essentially different in these
two cases.

\subsection{The case of integer $\sigma \leq 0$}
\label{pp_A}

For a nonpositive integer $\sigma$, the normalized coefficients
$C_{n}(z,\sigma)$ are
	\begin{equation}
C^{(-)}_{n}(z,\sigma) = \left\{
 \begin{array}{c}
{\displaystyle
 \frac{1}{\sqrt{I_{0}(2|z|)}}
      \frac{z^{n-|\sigma|}}{(n-|\sigma|)!} , }
  \;\;\;   n \geq |\sigma| , \\
 0, \;\;\; n < |\sigma|  .  \end{array}  \right.
\label{4.6}
	\end{equation}
The index `$-$' stands here and in the following for negative values
of $\sigma$, though all results are valid also for $\sigma = 0$.
It is interesting to note that the states
	\begin{equation}
|z,\sigma\rangle_{-} = \frac{1}{\sqrt{I_{0}(2|z|)}}
\sum_{n=|\sigma|}^{\infty} \frac{z^{n-|\sigma|}}{(n-|\sigma|)!}
|n\rangle
\label{4.7}
	\end{equation}
are not only the eigenstates of the $\hat{Z}(\sigma)$ with
eigenvalues $z$,
but also the eigenstates of the operator
	\begin{equation}
\hat{\Sigma}(z) \equiv \hat{n} - z\widehat{e^{-i\phi}}
\label{4.8}
	\end{equation}
with eigenvalues $|\sigma| = -\sigma$,
	\begin{equation}
\hat{\Sigma}(z) |z,\sigma\rangle_{-} = -
\sigma |z,\sigma\rangle_{-} .
\label{4.9}
	\end{equation}
We deduce from this equation the following recursion relation:
	\begin{equation}
C^{(-)}_{n+1}(z,\sigma) =\frac{z}{n+\sigma+1} C^{(-)}_{n}(z,\sigma) ,
\mbox{\hspace{1cm}} n \geq -1 , \;\; C^{(-)}_{-1}(z,\sigma) = 0 .
\label{4.10}
	\end{equation}
The difference between recursion relations (\ref{4.4}) and
(\ref{4.10}) is very essential, since Eq.\ (\ref{4.4}) is meaningful
for arbitrary values of $\sigma$, while Eq.\ (\ref{4.10}) has a
nonzero
solution only for a nonpositive integer $\sigma$.

Equations (\ref{4.9}) and (\ref{4.10}) can be transformed into
the differential equation \cite{BBA_PR2}:
	\begin{equation}
\left[ i\frac{d}{d\theta} +\sigma - z e^{-i\theta} \right]
\Theta^{(-)}(z,\sigma;\theta) = 0 .
\label{4.11}
	\end{equation}
Here, by the usual notation, $\Theta^{(-)}(z,\sigma;\theta)$ is the
phase representation function for the states $|z,\sigma\rangle_{-}$:
	\begin{equation}
\Theta^{(-)}(z,\sigma;\theta) = \sqrt{2\pi}\,
\langle\theta|z,\sigma\rangle_{-}
= \sum_{n = 0}^{\infty} C^{(-)}_{n}(z,\sigma) e^{-in\theta} ,
\label{4.12}
	\end{equation}
	\begin{equation}
|z,\sigma\rangle_{-} = \frac{1}{\sqrt{2\pi}}
\int_{\theta_{0}}^{\theta_{0}+ 2\pi } d\theta \,
\Theta^{(-)}(z,\sigma;\theta) |\theta\rangle .
\label{4.13}
	\end{equation}
The normalized solution of Eq.\ (\ref{4.11}) is
	\begin{equation}
\Theta^{(-)}(z,\sigma;\theta) = \frac{
\exp \left( i\sigma\theta + z e^{-i\theta} \right)
}{\sqrt{I_{0}(2|z|)}} ,
\label{4.14}
	\end{equation}
The $\Theta^{(-)}(z,\sigma;\theta)$ function must be $2\pi$ periodic.
Therefore $\sigma$ must be an integer. Moreover,
from Eq.\ (\ref{4.12}) we see that the
$\Theta^{(-)}(z,\sigma;\theta)$
function written as Fourier series must have only Fourier
coefficients with non-negative values of $n$ (this is a general
requirement to the phase representation functions \cite{BBA_PR2}).
For the function (\ref{4.14})
this demand forbids positive values of $\sigma$.
{}From Eq.\ (\ref{4.14}) we see that the states $|z,\sigma\rangle_{-}$
are philophase states, and the BG subcoherent
states $|\frac{1}{2},z\rangle$ are a special case corresponding to
$\sigma = 0$. The $|z,\sigma\rangle_{-}$ states with different
values of $\sigma$ have
the same phase properties because the phase distribution function
	\begin{equation}
Q^{(-)}(z;\theta) = \frac{1}{2\pi}
|\Theta^{(-)}(z,\sigma;\theta)|^{2}
= \frac{\exp [2|z|\cos(\theta-\bar{\varphi}_{z})]
}{2\pi I_{0}(2|z|)} , \mbox{\hspace{0.8cm}}
\bar{\varphi}_{z} = {\rm arg}\, z
\label{4.15}
	\end{equation}
does not depend on $\sigma$. This function was considered in
Sec.\ \ref{bg_D}. There we have discussed
in detail the phase properties of the states
$|k=\frac{1}{2},z\rangle = |z,\sigma=0\rangle$,
and this discussion remains unchanged with regard to all the states
$|z,\sigma\rangle_{-}$.

The statistical properties of the philophase states
$|z,\sigma\rangle_{-}$
can be calculated by using the photon-number distribution
	\begin{equation}
P^{(-)}_{n}(z,\sigma) = |\langle n|z,\sigma\rangle_{-}|^{2} =
\left\{ \begin{array}{c}   {\displaystyle
\frac{|z|^{2(n-|\sigma|)}}{I_{0}(2|z|)[(n-|\sigma|)!]^{2}} , }
\;\;\;    n \geq |\sigma| , \\
 0, \;\;\; n < |\sigma|  .  \end{array}  \right.
\label{4.16}
	\end{equation}
Another way is associated with the phase-state representation. For a
pure state $|f\rangle$ with the phase representation function
$\Theta(f;\theta)$, the number-operator moments are given by
\cite{BBA_PR2}
	\begin{equation}
\langle f|\hat{n}^{p}|f \rangle = \frac{i^{p}}{2\pi}
\int_{\theta_{0}}^{\theta_{0}+ 2\pi } d\theta \,
\Theta^{\ast}(f;\theta) \frac{d^{p}}{d\theta^{p}} \Theta(f;\theta) .
\label{4.17}
	\end{equation}
In either way, the calculation is simple, and we obtain
	\begin{eqnarray}
\langle \hat{n} \rangle^{(-)}_{z,\sigma} & = &
|z| \frac{I_{1}(2|z|)}{I_{0}(2|z|)} - \sigma ,
\label{4.18} \\
\langle \hat{n}^{2} \rangle^{(-)}_{z,\sigma} & = & \sigma^{2} +
|z|^{2} - 2\sigma |z|
\frac{I_{1}(2|z|)}{I_{0}(2|z|)} .
\label{4.19}
	\end{eqnarray}
The number variance,
	\begin{equation}
(\Delta n)^{2}_{z}(-) = |z|^{2} \left[ 1 -
\frac{I_{1}^{2}(2|z|)}{I_{0}^{2}(2|z|)} \right] ,
\label{4.20}
	\end{equation}
is independent of $\sigma$, just as the $Q^{(-)}(z;\theta)$ function,
and therefore the number-phase uncertainty relations for the
philophase states $|z,\sigma\rangle_{-}$ are the same as those ones
for the BG states $|k=\frac{1}{2},z\rangle = |z,\sigma=0\rangle$.
We have discussed these number-phase uncertainty relations in
Sec.\ \ref{bg_D}. The intensity correlation function is given by
	\begin{equation}
g^{(2)}_{z,\sigma}(-) = \left[ \sigma^{2} + |z|^{2} -
(2\sigma + 1)|z|
\frac{I_{1}^{2}(2|z|)}{I_{0}^{2}(2|z|)} + \sigma \right]
\left[ \sigma^{2} -
2\sigma |z| \frac{I_{1}^{2}(2|z|)}{I_{0}^{2}(2|z|)} +
|z|^{2} \frac{I_{1}^{2}(2|z|)}{I_{0}^{2}(2|z|)} \right]^{-1} .
\label{4.21}
	\end{equation}
The $g^{(2)}_{z,\sigma}(-)$ is always less than unity,
so the photon number distribution $P^{(-)}_{n}(z,\sigma)$ of
the philophase states $|z,\sigma\rangle_{-}$ is sub-Poissonian.
For $|z| \ll 1$, we get
	\begin{eqnarray}
\langle \hat{n} \rangle^{(-)}_{z,\sigma} & \approx &
|z|^{2} - |z|^{4}/2 - \sigma ,
\label{4.22} \\
\langle \hat{n}^{2} \rangle^{(-)}_{z,\sigma} & \approx &
\sigma^{2} + (1-2\sigma)|z|^{2} + \sigma |z|^{4},
\label{4.23} \\
g^{(2)}_{z,\sigma}(-) & \approx & 1 - \frac{|z|^{4}/2 -\sigma
}{ \sigma^{2}- 2\sigma (|z|^{2}-|z|^{4}/2) + |z|^{4} } .
\label{4.24}
	\end{eqnarray}
For $\sigma = 0$, we obtain $g^{(2)}_{z,0} \approx \frac{1}{2}$,
in accordance with Eq.\ (\ref{3.18}). When $|\sigma| \geq 1$,
expression (\ref{4.24}) can be further approximated:
	\begin{equation}
g^{(2)}_{z,\sigma}(-) \approx 1 - \frac{1}{|\sigma|} .
\label{4.25}
	\end{equation}
This result is interesting since, for $\sigma = -1$, the intensity
correlation function tends to its minimal allowed value:
$g^{(2)}_{z,-1}(-) \approx 0$. This is the maximal accessible
antibunching. For $\sigma = -2$, the $g^{(2)}_{z,\sigma}(-)$ again
approaches $\frac{1}{2}$, as for $\sigma = 0$. With further increase
of $|\sigma|$,
the $g^{(2)}_{z,\sigma}(-)$ tends to the Poissonian value $1$.
The intensity correlation function (\ref{4.25}) is the same as
that for the number states $|n\rangle$ with $n = |\sigma|$.
It is not surprising, because we see from Eq. (\ref{4.7}) that the
philophase states $|z,\sigma\rangle_{-}$ tend in the limit
$|z| \rightarrow 0$ to the number states $|n = |\sigma|\rangle$.
For $|z| \gg 1$, we get
	\begin{eqnarray}
\langle \hat{n} \rangle^{(-)}_{z,\sigma} & \approx &
|z| - \sigma - \mbox{{\small $\frac{1}{4}$}} ,
\label{4.26} \\
\langle \hat{n}^{2} \rangle^{(-)}_{z,\sigma} & \approx &
\sigma^{2} + |z|^{2} - 2\sigma(|z|-\mbox{{\small $\frac{1}{4}$}}) ,
\label{4.27} \\
g^{(2)}_{z,\sigma}(-) & \approx &
1 - \frac{|z|/2}{\sigma^{2} + |z|^{2} - 2\sigma (|z|-\mbox{{\small
$\frac{1}{4}$}}) -|z|/2} .
\label{4.28}
	\end{eqnarray}
When $|z| \gg |\sigma|$, expression (\ref{4.28}) can be further
approximated:
	\begin{equation}
g^{(2)}_{z,\sigma}(-) \approx 1 - \frac{1}{2|z|} .
\label{4.29}
	\end{equation}
All the BG states in the classical limit behave according to Eq.\
(\ref{4.29}), and this behavior also holds for the eigenstates
of the generalized operator $\hat{Z}(\sigma)$.

We finish the discussion of the case $\sigma \leq 0$ by noting that
the philophase states $|z,\sigma\rangle_{-}$ arise from the problem
of finding intelligent states for operators which are combinations
of the number and phase-related operators \cite{BBA_PR2}.
We define two Hermitian operators:
	\begin{eqnarray}
\hat{\Sigma}_{1}(z) & = & \hat{n} - ({\rm Re}\, z)
\hat{C}
- ({\rm Im}\, z) \hat{S} ,
\label{4.30} \\
\hat{\Sigma}_{2}(z) & = & ({\rm Re}\, z) \hat{S}
- ({\rm Im}\, z) \hat{C} ,
\label{4.31}
	\end{eqnarray}
such that
	\begin{equation}
\hat{\Sigma}(z) = \hat{\Sigma}_{1}(z) + i\hat{\Sigma}_{2}(z) .
\label{4.32}
	\end{equation}
Therefore, the eigenvalue equation (\ref{4.9}) means that the
states $|z,\sigma\rangle_{-}$ are intelligent states for operators
$\hat{\Sigma}_{1}(z)$ and $\hat{\Sigma}_{2}(z)$, i.e., for these
states the uncertainty relation
	\begin{equation}
[ {}^{\ast}_{\ast} ( \Delta \hat{\Sigma}_{1} )^{2} {}^{\ast}_{\ast} ]
[ {}^{\ast}_{\ast} ( \Delta \hat{\Sigma}_{2} )^{2} {}^{\ast}_{\ast} ]
\geq \frac{1}{4} |\langle {}^{\ast}_{\ast} [\hat{\Sigma}_{1},
\hat{\Sigma}_{2}] {}^{\ast}_{\ast} \rangle |^{2}
\label{4.33}
	\end{equation}
is an equality. Indeed, we can use results of this and preceding
sections [Eqs.\ (\ref{3.46}), (\ref{3.47}) and
(\ref{4.18})--(\ref{4.20})] in order to calculate
	\begin{equation}
{}^{\ast}_{\ast} \! \left( \Delta \hat{\Sigma}_{1}
\right)^{2}_{z,\sigma}\!(-) {}^{\ast}_{\ast}
= {}^{\ast}_{\ast} \! \left( \Delta \hat{\Sigma}_{2}
\right)^{2}_{z,\sigma}\!(-) {}^{\ast}_{\ast}
= \frac{|z|}{2} \frac{I_{1}(2|z|)}{I_{0}(2|z|)} .
\label{4.34}
	\end{equation}
The commutation relation is
	\begin{equation}
{}^{\ast}_{\ast} [\hat{\Sigma}_{1},\hat{\Sigma}_{2}]
{}^{\ast}_{\ast} =
i [ ({\rm Re}\, z) \hat{C} + ({\rm Im}\, z) \hat{S} ] .
\label{4.35}
	\end{equation}
The expectation value is easily obtained by using Eq.\ (\ref{3.46}):
	\begin{equation}
\langle {}^{\ast}_{\ast} [\hat{\Sigma}_{1},\hat{\Sigma}_{2}]
{}^{\ast}_{\ast}
\rangle^{(-)}_{z,\sigma} = i|z| \frac{I_{1}(2|z|)}{I_{0}(2|z|)} ,
\label{4.36}
	\end{equation}
and it is evident now that the states $|z,\sigma\rangle_{-}$
are the intelligent states for operators $\hat{\Sigma}_{1}(z)$ and
$\hat{\Sigma}_{2}(z)$.

\subsection{The case of integer $\sigma \geq 0$}
\label{pp_B}

As soon as $\sigma$ passes through zero, the properties of the states
$|z,\sigma\rangle$ sharply change. For $\sigma \geq 0$ these states
are
	\begin{equation}
|z,\sigma\rangle_{+} = \frac{1}{\sqrt{T_{0}(|z|,\sigma)}}
\sum_{n = 0}^{\infty}
\frac{z^{n+\sigma}}{(n+\sigma)!} |n\rangle ,
\label{4b1}
	\end{equation}
where we have defined
	\begin{equation}
T_{0}(|z|,\sigma) \equiv I_{0}(2|z|) - \sum_{m=0}^{\sigma-1}
\frac{|z|^{2m}}{(m!)^{2}} .
\label{4b2}
	\end{equation}
The index `$+$' stands here and in the following for positive values
of $\sigma$, though all results are valid also for $\sigma = 0$.
The states $|z,\sigma\rangle_{+}$ are normalized but not orthogonal
to each other:
	\begin{equation}
{}_{+}\langle z_{1},\sigma|z_{2},\sigma\rangle_{+}
= \frac{ T_{0}(\sqrt{z_{1}^{\ast}z_{2}},\sigma)
}{ \sqrt{ T_{0}(|z_{1}|,\sigma) T_{0}(|z_{2}|,\sigma) } } .
\label{4b3}
	\end{equation}
In the following we will consider a modified HP realization
of the SU(1,1) Lie algebra. After that we will discuss statistical
and phase properties of the $|z,\sigma\rangle_{+}$ states and their
relation to the Glauber CS.

\subsubsection{Modified HP realization}

It is easy to prove that the states $|z,\sigma\rangle_{+}$ resolve
the identity:
	\begin{equation}
\int d\mu(z,\sigma) |z,\sigma\rangle_{+} {}_{+}\langle z,\sigma|
=\hat{1} ,
\label{4b4}
	\end{equation}
where
	\begin{equation}
d\mu(z,\sigma) = \frac{2}{\pi} K_{0}(2|z|)
T_{0}(|z|,\sigma) d^{2}\! z .
\label{4b5}
	\end{equation}
Indeed, Eq.\ (\ref{4b4}) reads
	\begin{equation}
4 \sum_{n = 0}^{\infty}
\frac{|n\rangle\langle n|}{[(n+\sigma)!]^{2}} \int_{0}^{\infty}
d|z|\, K_{0}(2|z|) |z|^{2n+2\sigma+1} = \hat{1} .
\label{4b6}
	\end{equation}
By using formula (\ref{3.6}), we make sure of this equality.
Thus the states $|z,\sigma\rangle_{+}$ form, for each integer
$\sigma \geq 0$,
an overcomplete basis in the harmonic oscillator Hilbert space.
This result leads us to idea that the operator $\hat{Z}(\sigma)$,
whose eigenstates the $|z,\sigma\rangle_{+}$ states are, is the
lowering generator belonging to a realization of the SU(1,1) Lie
algebra. This operator is a
generalization of the lowering generator
$\hat{K}_{-}(k=\frac{1}{2}) =
\hat{Z}(\sigma=0)$, and therefore a modified realization with
$\sigma \geq 0$ should be a generalization of the HP realization
with $k=\frac{1}{2}$. We introduce the following sets of operators:
	\begin{equation}
\begin{array}{rcl}
\hat{K}_{-}(\sigma) & = & \widehat{e^{i\phi}} (\hat{n} +\sigma)
= \hat{Z}(\sigma) , \\
\hat{K}_{+}(\sigma) & = & (\hat{n} +\sigma) \widehat{e^{-i\phi}}
= [\hat{K}_{-}(\sigma)]^{\dagger} , \\
\hat{K}_{3}(\sigma) & = & \hat{n} +\sigma + \frac{1}{2} .
\end{array}
\label{4b7}
	\end{equation}
These operators obey the SU(1,1) Lie algebra provided the antinormal
ordering is used:
	\begin{equation}
{}^{\ast}_{\ast} [\hat{K}_{-}(\sigma),\hat{K}_{+}(\sigma)]
{}^{\ast}_{\ast} = 2 \hat{K}_{3}(\sigma) ,
\;\;\;\;\;
[\hat{K}_{3}(\sigma),\hat{K}_{\pm}(\sigma)] =
\pm \hat{K}_{\pm}(\sigma) .
\label{4b8}
	\end{equation}
[The subscripts `$-$' and `$+$' of the SU(1,1) generators are not
related to the sign of $\sigma$.]
The antinormal ordering should be applied also to the calculation of
the Casimir operator:
	\begin{equation}
\hat{Q} = [\hat{K}_{3}(\sigma)]^{2} - {}^{\ast}_{\ast} \frac{1}{2}[
\hat{K}_{+}(\sigma) \hat{K}_{-}(\sigma)
+ \hat{K}_{-}(\sigma) \hat{K}_{+}(\sigma) ] {}^{\ast}_{\ast} = -
\frac{1}{4} \hat{1} .
\label{4b9}
	\end{equation}
By comparing this result with Eq.\ (\ref{2.5}), we see that  the
modified HP realization (\ref{4b7}) corresponds to the
case of the discrete series representation with $k=\frac{1}{2}$.
The action
of the generators (\ref{4b7}) on the number states is given by
	\begin{equation}
\begin{array}{rcl}
\hat{K}_{3}(\sigma) |n\rangle & = & (n+\sigma+\mbox{{\small
$\frac{1}{2}$}}) |n\rangle , \\
\hat{K}_{+}(\sigma) |n\rangle & = & (n+\sigma+1) |n+1\rangle , \\
\hat{K}_{-}(\sigma) |n\rangle & = & (n+\sigma) |n-1\rangle .
\end{array}
\label{4b10}
	\end{equation}
It follows from the comparison of these formulas with relations
(\ref{2.7}) that in the present modification
the orthonormal basis $|k=\frac{1}{2},m\rangle$ of the
discrete series state space is somewhat different from the
usually used one. The modified orthonormal basis is given by
	\begin{equation}
|k=\mbox{{\small $\frac{1}{2}$}},m\rangle = |n\rangle ,
\mbox{\hspace{1cm}} m=n+\sigma ,
\label{4b11}
	\end{equation}
where $|n\rangle$ $(n=0,1,\ldots,\infty)$ is the number-state basis,
so that $m=\sigma,\sigma+1,\ldots,\infty$. Then the completeness
relation is
	\begin{equation}
\sum_{m=\sigma}^{\infty} |\mbox{{\small $\frac{1}{2}$}},m\rangle
\langle\mbox{{\small $\frac{1}{2}$}},m| = \hat{1} .
\label{4b12}
	\end{equation}
In the modified HP realization (\ref{4b7}) the index $m$
goes from $\sigma$ and not from zero, as it is customary.
Thus we get, for $\sigma > 0$, a generalization of the usual SU(1,1)
discrete series representation.

By using the identity resolution (\ref{4b4}), we construct the
Hilbert space of entire functions $f(z,\sigma)$, which are analytic
over the whole $z$ plane. For a normalized state $|f\rangle$ of the
form (\ref{2.15}), we get
	\begin{equation}
f(z,\sigma) = \sqrt{T_{0}(|z|,\sigma)} \,
{}_{+}\!\langle z^{\ast},\sigma|f\rangle =
\sum_{n = 0}^{\infty} C_{n}(f) \frac{z^{n+\sigma}}{(n+\sigma)!} ,
\label{4b14}
	\end{equation}
and this state can be represented in the $|z,\sigma\rangle_{+}$
basis:
	\begin{equation}
|f\rangle = \int d\mu(z,\sigma)
\frac{f(z^{\ast},\sigma)}{\sqrt{T_{0}(|z|,\sigma)}}
 |z,\sigma\rangle_{+} .
\label{4b15}
	\end{equation}
The orthonormal basis $u_{n}(z,\sigma)$ in the Hilbert space of
entire functions can be chosen corresponding to the number-state
basis:
	\begin{equation}
u_{n}(z,\sigma) = \sqrt{T_{0}(|z|,\sigma)} \,
{}_{+}\!\langle z^{\ast},\sigma|n\rangle =
\frac{z^{n+\sigma}}{(n+\sigma)!} .
\label{4b16}
	\end{equation}
The generators $\hat{K}_{\pm}(\sigma)$ and $\hat{K}_{3}(\sigma)$
act on the Hilbert space of entire functions $f(z,\sigma)$ as
linear operators:
	\begin{equation}
\hat{K}_{+}(\sigma) = z, \;\;\;\;\;
\hat{K}_{-}(\sigma) = \frac{d}{dz} + z \frac{d^{2}}{dz^{2}} ,
\;\;\;\;\;
\hat{K}_{+}(\sigma) = z \frac{d}{dz} + \frac{1}{2} .
\label{4b17}
	\end{equation}
We obtain, for example,
	\begin{equation}
\begin{array}{rcl}
\hat{K}_{3}(\sigma) u_{n}(z,\sigma) & = &
(n +\sigma + \mbox{{\small $\frac{1}{2}$}}) u_{n}(z,\sigma) , \\
\hat{K}_{+}(\sigma) u_{n}(z,\sigma) & = &
(n +\sigma + 1) u_{n+1}(z,\sigma) , \\
\hat{K}_{-}(\sigma) u_{n}(z,\sigma) & = &
(n +\sigma) u_{n-1}(z,\sigma) .
\end{array}
\label{4b18}
	\end{equation}
We see that the unusual features of the modified HP
realization manifest in the fact that the vacuum property,
	\begin{equation}
\hat{K}_{-}(\sigma) u_{0}(z,\sigma) = 0 ,
\label{4b19}
	\end{equation}
must be included as an additional restriction. This fact
is related to the use of the antinormal ordering, that is
introduced as an additional restriction in order to restore
the unitarity of the exponential phase operators.

\subsubsection{Statistical and phase properties}

The photon-number distribution of the $|z,\sigma\rangle_{+}$
states is
	\begin{equation}
P^{(+)}_{n}(z,\sigma) = |\langle n|z,\sigma\rangle_{+}|^{2} =
\frac{1}{T_{0}(|z|,\sigma)}
\frac{|z|^{2(n+\sigma)}}{[(n+\sigma)!]^{2}} .
\label{4b20}
	\end{equation}
After some algebra we obtain
	\begin{eqnarray}
\langle \hat{n} \rangle^{(+)}_{z,\sigma} & = &
|z| \frac{T_{1}(|z|,\sigma)}{T_{0}(|z|,\sigma)} - \sigma ,
\label{4b21} \\
\langle \hat{n}^{2} \rangle^{(+)}_{z,\sigma} & = & \sigma^{2}
+ |z|^{2} - 2\sigma |z|
\frac{T_{1}(|z|,\sigma)}{T_{0}(|z|,\sigma)} +
\frac{1}{T_{0}(|z|,\sigma)}
\frac{|z|^{2\sigma}}{[\Gamma(\sigma)]^{2}} ,
\label{4b22}
	\end{eqnarray}
where we have defined
	\begin{equation}
T_{1}(|z|,\sigma) \equiv I_{1}(2|z|) - \sum_{m=0}^{\sigma-2}
\frac{|z|^{2m+1}}{m!(m+1)!} ,
\label{4b23}
	\end{equation}
and the $T_{0}(|z|,\sigma)$ is defined by Eq.\ (\ref{4b2}).
The intensity correlation function is
	\begin{eqnarray}
g^{(2)}_{z,\sigma}(+) & = & \left[ \sigma^{2} +|z|^{2}
+ \sigma - (2\sigma+1)|z|
\frac{T_{1}(|z|,\sigma)}{T_{0}(|z|,\sigma)} +
\frac{1}{T_{0}(|z|,\sigma)}
\frac{|z|^{2\sigma}}{[\Gamma(\sigma)]^{2}}
\right] \nonumber \\
& & \times \left[ \sigma^{2} + |z|^{2}
\frac{T_{1}^{2}(|z|,\sigma)}{T_{0}^{2}(|z|,\sigma)}
- 2\sigma|z| \frac{T_{1}(|z|,\sigma)}{T_{0}(|z|,\sigma)}
\right]^{-1} .
\label{4b24}
	\end{eqnarray}
Only for $\sigma=0$ and $\sigma=1$ the
photon statistics is sub-Poissonian for all values of $|z|$, i.e.,
the $g^{(2)}_{z,\sigma}(+)$ is less than unity in the whole $z$ plane.
For $\sigma \geq 2$, the $g^{(2)}_{z,\sigma}(+)$ is greater than
unity while $|z|$ is less than a specific value depending on $\sigma$.
When $|z|$ exceeds this value, the intensity correlation function
becomes less than unity. In the quantum limit $|z| \ll 1$, we get
	\begin{eqnarray}
\langle \hat{n} \rangle^{(+)}_{z,\sigma} & \approx &
\frac{|z|^{2}}{(\sigma+1)^{2}} +
\frac{(\sigma^{2}-2) |z|^{4}}{(\sigma+1)^{4}(\sigma+2)^{2}} ,
\label{4b25} \\
\langle \hat{n}^{2} \rangle^{(+)}_{z,\sigma} & \approx &
\frac{|z|^{2}}{(\sigma+1)^{2}}  +
\frac{(3\sigma^{2}+4\sigma) |z|^{4}}{(\sigma+1)^{4}(\sigma+2)^{2}} ,
\label{4b26} \\
g^{(2)}_{z,\sigma}(+) & \approx &
2 \left( \frac{\sigma+1}{\sigma+2} \right)^{2} .
\label{4b27}
	\end{eqnarray}
For $\sigma=0$, we return to the known value $\frac{1}{2}$; for
$\sigma=1$,
the $g^{(2)}_{z,\sigma}(+)$ tends to $\frac{8}{9}$; for
$\sigma \geq 2$, the
limiting values of the $g^{(2)}_{z,\sigma}(+)$ are greater than
unity. For very large values of $\sigma$ ($\sigma \gg 1$), the
intensity correlation function (\ref{4b27}) tends to the thermal
value $2$. In the limit $|z| \gg \sigma$, we get
	\begin{eqnarray}
\langle \hat{n} \rangle^{(+)}_{z,\sigma} & \approx &
|z| - \sigma -\frac{1}{4} ,
\label{4b28} \\
\langle \hat{n}^{2} \rangle^{(+)}_{z,\sigma} & \approx &
|z|^{2} + \sigma^{2} - 2\sigma (|z| - \mbox{{\small
$\frac{1}{4}$}}) ,
\label{4b29} \\
g^{(2)}_{z,\sigma}(+) & \approx &
1 - \frac{1}{2|z|} .
\label{4b30}
	\end{eqnarray}
Thus, the $|z,\sigma\rangle_{+}$ states
tend, in this limit, to have Poissonian photon statistics
in the same way as the BG subcoherent states
and the $|z,\sigma\rangle_{-}$ states.

The phase representation function of the $|z,\sigma\rangle_{+}$
states is
	\begin{equation}
\Theta^{(+)}(z,\sigma;\theta) =
\frac{e^{i\sigma\theta}}{\sqrt{T_{0}(|z|,\sigma)}}
\left[ \exp \left( z e^{-i\theta} \right) - \sum_{m=0}^{\sigma-1}
\frac{ \left( z e^{-i\theta} \right)^{m} }{ m! } \right] .
\label{4b31}
	\end{equation}
When $\sigma$ is of order of unity, we can refer to the
$|z,\sigma\rangle_{+}$ states as philophase states. However, when
$\sigma$
increases, more and more additional terms are included in the
$\Theta^{(+)}(z,\sigma;\theta)$ function, so that the
$|z,\sigma\rangle_{+}$ states
become unsuitable to be called philophase states.
The phase distribution function is given by
	\begin{eqnarray}
Q^{(+)}(z,\sigma;\theta) & = &
\frac{1}{2\pi} |\Theta^{(+)}(z,\sigma;\theta)|^{2} \nonumber \\
& = & \frac{1}{2\pi T_{0}(|z|,\sigma)} \left\{ \exp \left[
2|z| \cos (\theta-\bar{\varphi}_{z}) \right] +
\sum_{n,m=0}^{\sigma-1}
\frac{|z|^{n+m}}{n!m!} \exp \left[i (n-m) (\theta-\bar{\varphi}_{z})
\right] \right. \nonumber \\
&  & \left.
- \exp \left( z e^{-i\theta} \right) \sum_{m=0}^{\sigma-1}
\frac{ \left( z^{\ast} e^{i\theta} \right)^{m} }{ m! }
- \exp \left( z^{\ast} e^{i\theta} \right) \sum_{m=0}^{\sigma-1}
\frac{ \left( z e^{-i\theta} \right)^{m} }{ m! } \right\} .
\label{4b32}
	\end{eqnarray}
The larger values of
$\sigma$ are, the flatter the $Q^{(+)}(z,\sigma;\theta)$ function
is, and
the worse the phase of the state is defined. By using the phase
distribution (\ref{4b32}), we calculate numerically the phase
variance ${}^{\ast}_{\ast}(\Delta\phi)^{2}_{z,\sigma}(+)
{}^{\ast}_{\ast}$.
For $|z| \rightarrow 0$, the
${}^{\ast}_{\ast}(\Delta\phi)^{2}_{z,\sigma}(+){}^{\ast}_{\ast}$
tends to the random value $\pi^{2}/3$. For $|z| \gg \sigma$, the
phase variance tends to zero.
For given $|z|$, the larger values of $\sigma$ are, the more
uncertain the phase of the state is.

We study the number-phase uncertainty relation of the
$|z,\sigma\rangle_{+}$ states by calculating numerically the
${\cal V}^{(+)}(z,\sigma)$ function, defined according to the
general expression (\ref{2.81}).
When $|z| \rightarrow 0$, the ${\cal V}^{(+)}(z,\sigma)$ function
tends, for any $\sigma$, to the standard random-phase value
$\pi^{2}/12$. While $|z|$ increases, the ${\cal V}^{(+)}(z,\sigma)$
function at first grows, reaches a maximum at a specific value
of $|z|$ depending on $\sigma$, and then decreases tending, for
$|z| \gg \sigma$, to the limiting value $\frac{1}{4}$.
For $\sigma=0$, the
maximum is already at  $|z|=0$. The larger values of $\sigma$ are,
the higher the maximum of the ${\cal V}^{(+)}(z,\sigma)$ function is,
and the slower an equality is achieved, as $|z|$ increases, in the
number-phase uncertainty relation.

\subsubsection{Contraction to the Glauber CS}

Consider the boson annihilation operator
$\hat{a} = \widehat{e^{i\phi}}
\sqrt{\hat{n}}$ acting on a photon state with mean photon number
$\langle\hat{n}\rangle$ and photon-number variance $(\Delta n)^{2}$.
Defining $\delta\hat{n} \equiv \hat{n} - \langle\hat{n}\rangle$,
we can write
$\hat{n} = \langle\hat{n}\rangle + \delta\hat{n}$. We choose a
state for which
$\Delta n \ll \langle\hat{n}\rangle$. Then we can approximate the
annihilation operator $\hat{a}$ acting on such a state:
	\begin{equation}
\hat{a} = \widehat{e^{i\phi}} \sqrt{\langle\hat{n}\rangle +
\delta\hat{n}} \approx
\widehat{e^{i\phi}} \sqrt{\langle\hat{n}\rangle} \left( 1 +
\frac{\delta\hat{n}}{2\langle\hat{n}\rangle}
\right) = \frac{1}{2\sqrt{\langle\hat{n}\rangle}}
\widehat{e^{i\phi}} (\hat{n} + \langle\hat{n}\rangle) .
\label{4b33}
	\end{equation}
We see that in the described case the annihilation operator
$\hat{a}$
can be approximated, up to a numerical factor, by the
operator $\hat{Z}(\sigma) = \widehat{e^{i\phi}} (\hat{n} + \sigma)$,
	\begin{equation}
\hat{a} \approx \frac{1}{2\sqrt{\sigma}} \hat{Z}(\sigma) ,
\mbox{\hspace{1cm}} \sigma = \langle\hat{n}\rangle ,
\;\; \Delta n \ll \langle\hat{n}\rangle .
\label{4b34}
	\end{equation}
Therefore, the $|z,\sigma\rangle_{+}$ eigenstates of the operator
$\hat{Z}(\sigma)$ are, in the considered case, an approximation of
the Glauber CS $|\alpha\rangle$, which are the eigenstates of the
$\hat{a}$,
	\begin{equation}
|z,\sigma\rangle_{+} \approx |\alpha\rangle , \mbox{\hspace{1cm}}
\Delta n \ll \langle\hat{n}\rangle .
\label{4b35}
	\end{equation}
By comparing eigenvalues, we get
	\begin{equation}
\alpha \approx \frac{z}{2\sqrt{\sigma}} .
\label{4b36}
	\end{equation}
The mean photon number of the Glauber CS is
$\langle\hat{n}\rangle_{\alpha}
= |\alpha|^{2}$. We suppose that in the considered case the mean
photon number $\langle \hat{n} \rangle^{(+)}_{z,\sigma}$ of the
$|z,\sigma\rangle_{+}$ states is approximately equal to the
$\langle\hat{n}\rangle_{\alpha}$, and this gives $\sigma =
\langle \hat{n} \rangle^{(+)}_{z,\sigma} \approx |\alpha|^{2}$.
Then we find from Eq.\ (\ref{4b36})
	\begin{equation}
|z| \approx 2|\alpha|^{2} \approx 2\sigma .
\label{4b37}
	\end{equation}
The number variance of the Glauber CS is
$(\Delta n)^{2}_{\alpha} = |\alpha|^{2}$. Then the condition
$\Delta n \ll \langle\hat{n}\rangle$ is satisfied for
$|\alpha| \gg 1$.
The conclusion is that the $|z,\sigma\rangle_{+}$ states contract to
the Glauber CS $|\alpha\rangle$ for $\sigma \approx |z|/2 \approx
|\alpha|^{2}$ provided that $|\alpha| \gg 1$ (the classical
limit).

This result can be verified by calculating statistical properties
of the $|z,\sigma\rangle_{+}$ states with $|z| = 2\sigma$.
For large values
of $\sigma$, we would get, according to the discussed contraction,
	\begin{equation}
\langle \hat{n} \rangle^{(+)}_{z,\sigma} \approx
(\Delta n)^{2}_{z,\sigma}(+)
\approx \sigma .
\label{4b38}
	\end{equation}
This is confirmed by numerical calculations. By taking
$|z| = 2\sigma$ in Eqs.\ (\ref{4b21}) and (\ref{4b22}), we find that
the relations $\langle \hat{n} \rangle^{(+)}_{z,\sigma}/\sigma$ and
$(\Delta n)^{2}_{z,\sigma}(+)/\sigma$ quickly tend to unity, as
$\sigma$ increases. The difference from unity for
$\langle \hat{n} \rangle^{(+)}_{z,\sigma}/\sigma$
is about $5\times 10^{-3}$, as $\sigma$ goes to $50$, and for
$(\Delta n)^{2}_{z,\sigma}(+)/\sigma$ it is about $10^{-5}$, as
$\sigma$ goes to $20$. Also, we calculate numerically the intensity
correlation function $g^{(2)}_{z,\sigma}(+)$ of Eq.\ (\ref{4b24})
for $|z| = 2\sigma$. This function quickly tends to the Poissonian
value 1 of the Glauber CS. The difference of the
$g^{(2)}_{z,\sigma}(+)$ from unity  is about $5\times 10^{-4}$, as
$\sigma$ goes to $20$, and it is about $10^{-4}$, as
$\sigma$ goes to $50$.

\section{Discussion and conclusions}
\label{sec:conc}

The SU(1,1) CS $|k,\zeta\rangle$ and the BG subcoherent states
$|k,z\rangle$
can be considered as two possible modifications of the familiar
Glauber CS for the SU(1,1) Lie group. However, these two
modifications lead us to the states with very different
statistical and phase properties. The CS $|k,\zeta\rangle$
have wholly super-Poissonian statistics, while all the BG states
$|k,z\rangle$ are antibunched. With increase of the Bargmann index
$k$, the photon-number distributions of the $|k,\zeta\rangle$ and
$|k,z\rangle$ states move from opposite sides to the Poissonian
distribution.
Another difference between the $|k,\zeta\rangle$ and $|k,z\rangle$
states is the behavior of statistical properties with change of
the photon-excitation strength. As mean photon number increases,
the relative photon-number uncertainty
$\Delta n/\langle\hat{n}\rangle$
tends to zero for the BG states, but it approaches a nonzero limit
depending on $k$ for the CS $|k,\zeta\rangle$. Analogously, for
large excitations,
the intensity correlation function $g^{(2)}$ tends to unity
for the BG states, but it does not depend on the excitation
strength for the SU(1,1) CS.

Phase properties of the
$|k,\zeta\rangle$ and $|k,z\rangle$ states show opposite behaviors
with change of $k$. As $k$ increases, the phase distribution
$Q(\theta)$ becomes narrower for the SU(1,1) CS, and it becomes
flatter for the BG states. The behavior of the phase variance
${}^{\ast}_{\ast}(\Delta\phi)^{2}{}^{\ast}_{\ast}$ with change of
the photon-excitation
strength is similar for all discussed types of states. In the
quantum limit (near the vacuum) the phase variance tends to its
random value $\pi^{2}/3$, while in the classical limit
(large excitations) it approaches zero, as must be for a classical
wave with the perfectly defined phase. Phase properties of a photon
state can be studied by using the cosine or sine variances as well
as the variance of the Hermitian phase operator $\hat{\phi}$.
Results are equivalent for using any phase-related observable,
since optical phase is unique in the antinormal ordering formalism.

An important problem discussed in the present paper is the
uncertainty relation between the number and phase variables.
Again, the choice of phase-related observable for studying this
uncertainty relation is arbitrary. We have seen that by using
the phase variance
${}^{\ast}_{\ast}(\Delta\phi)^{2}{}^{\ast}_{\ast}$ and the cosine
variance ${}^{\ast}_{\ast}(\Delta C)^{2}{}^{\ast}_{\ast}$ we obtain
the same information.
The properties of the number-phase uncertainty relation are
essentially different for the $|k,\zeta\rangle$ and $|k,z\rangle$
states.
Near the vacuum ($|\zeta| \rightarrow 0$ or $|z| \rightarrow 0$)
the uncertainty function ${\cal V}$ tends to the random-phase value
$\pi^{2}/12$ for all types of states. However, in the limit of
large excitations the behaviors of the SU(1,1) CS and the BG states
are absolutely opposite. In this limit the ${\cal V}(k,\zeta)$
function of the CS $|k,\zeta\rangle$ blows up, according to the
specific statistical properties of these states, while the
${\cal V}(k,z)$ function of the BG states $|k,z\rangle$ tends to
its minimal possible value $\frac{1}{4}$. The larger values of $k$
are, the slower the $|k,z\rangle$ states bring an equality to the
number-phase uncertainty relation, as photon-excitation strength
increases. In contrast to that, the CS $|k,\zeta\rangle$ provide an
approximate equality in the number-phase uncertainty relation
for large values of $k$, in an intermediate range of excitations.
With all that, the BG states $|k,z\rangle$ have a simple
representation
in the coherent-state basis $|k,\zeta\rangle$, and vice versa.

The case of the Bargmann index $k$ equal to $\frac{1}{2}$ is
interesting from two points of view. Firstly, this case is related
to a simple type of intensity-dependent coupling in the
Jaynes-Cummings model Hamiltonians. Secondly, the photon states
associated with the HP SU(1,1) realization have, for
$k=\frac{1}{2}$, special phase
properties. Phase-state representation function $\Theta(\theta)$,
defined
generally as Fourier series, can be converted into a relatively
simple functional form for the philophase states
$|\frac{1}{2},\zeta\rangle$ and $|\frac{1}{2},z\rangle$.
Moreover, the SU(1,1) CS $|\frac{1}{2},\zeta\rangle$ also are
the eigenstates of the exponential phase operator
$\widehat{e^{i\phi}}$, just
like the phase states $|\theta\rangle$. From the other hand,
the phase properties of the BG states $|\frac{1}{2},z\rangle$ can be
generalized by introducing eigenstates $|z,\sigma\rangle$ of the
operator $\hat{Z}(\sigma) = \widehat{e^{i\phi}} (\hat{n} + \sigma)$.
For integer $\sigma \leq 0$, we find the class of generalized
philophase states $|z,\sigma\rangle_{-}$. All
of them have the same phase distribution function as the BG
states $|\frac{1}{2},z\rangle$.
The philophase states $|z,\sigma\rangle_{-}$
are antibunched, and in the quantum limit $|z| \rightarrow 0$
they tend to the number states $|n=|\sigma|\rangle$. In the case of
integer $\sigma \geq 0$, we find a modification of the HP SU(1,1)
realization with $k=\frac{1}{2}$. Statistical and phase properties
and the number-phase uncertainty relation of the states
$|z,\sigma\rangle_{+}$
have interesting features. For $\sigma$ of order of unity, the
$|z,\sigma\rangle_{+}$ states are close to be described as philophase
states, while for $\sigma \gg 1$, they contract to the Glauber CS
$|\alpha\rangle$, provided that $|z| \approx 2\sigma$ and
$\alpha \approx z/(2\sqrt{\sigma})$.

\acknowledgments
I thank S. Bespalko for fruitful discussions and versatile
help. This work could not be done without tight and fruitful
collaboration with Prof. Y. Ben-Aryeh. I am grateful to him
for helpful discussions and valuable ideas and remarks.
I thank A. Berengolts for great help in computer-concerned problems.
Extremely useful and informative discussions with Prof. J. Katriel
are acknowledged.

\begin{flushleft}

\end{flushleft}

\end{document}